\documentclass[review,12pt,sort&compress,times]{elsarticle}
\usepackage[utf8]{inputenc}
\usepackage{amsmath}
\usepackage{amsfonts}
\usepackage{amssymb}
\usepackage{amsthm}
\usepackage{lmodern}
\usepackage{graphicx}
\usepackage{empheq}
\usepackage{color}
\usepackage{subcaption}
\usepackage{tabularx}
\usepackage{upgreek}
\usepackage{natbib}
\usepackage[squaren,Gray]{SIunits}
\usepackage[left]{lineno}
\usepackage{physics} 

\newcommand{\vt}[1]{\mathbf{#1}}

\journal{Wave Motion}

\begin{document}

\begin{frontmatter}

\title{Forced vibrations and wave propagation in multilayered solid spheres using a one-dimensional semi-analytical finite element method}

\author[GeM,Ifsttar]{Matthieu Gallezot\corref{cor1}}
\ead{matthieu.gallezot@ifsttar.fr}
\cortext[cor1]{Corresponding author}
\address[GeM]{Université de Nantes, Institut de Recherche en Génie Civil et Mécanique, BP 92208, Nantes, France}

\author[Ifsttar]{Fabien Treyss\`{e}de}
\author[Ifsttar]{Odile Abraham}
\address[Ifsttar]{IFSTTAR, GERS, GeoEND, F-44344, Bouguenais, France}

\begin{abstract}
A numerical model is proposed to compute the eigenmodes and the forced response of multilayered elastic spheres. The main idea is to describe analytically the problem along the angular coordinates with spherical harmonics and to discretize the radial direction with one-dimensional finite elements. The proper test function must be carefully chosen so that both vector and tensor spherical harmonics orthogonality relationships can be used. The proposed approach yields a general one-dimensional formulation with a fully analytical description of the angular behaviour, suitable for any interpolating technique. A linear eigenvalue problem, simple and fast to solve, is then obtained. The eigensolutions are the spheroidal and torsional modes. They are favourably compared with literature results for a homogeneous sphere. The eigensolutions are superposed to compute explicitly the forced response. The latter is used to reconstruct the propagation of surfaces waves. In particular, the collimation of a Rayleigh wave (non-diffracted surface wave propagating with a quasi-constant width) excited by a line source in a homogeneous sphere is recovered with the model. Based on the vibration eigenmodes, a modal analysis shows that such a wave is a superposition of fundamental spheroidal modes with a displacement confined at the equator of the sphere. These modes are the so-called Rayleigh modes, of sectoral type and high polar wavenumbers. When a thin viscoelastic coating is added to the sphere, the Rayleigh mode behaviour is recovered in a limited frequency range, allowing the generation of a collimating wave at the interface between the sphere and the coating.
\end{abstract}


\end{frontmatter}

\section{Introduction}


The study of the free vibrations (eigenmodes) of an elastic sphere is a classical mechanics problem formally solved first by \citet{Lamb81}. This topic received a great interest in geophysics \citep{Sezawa27, Sato62}, using the Earth's eigenmodes to synthesize seismograms and improve the understanding of earthquakes \citep{Aki80}. \citet{Lamb81} and \citet{Shah69} also studied the case of a hollow sphere (spherical shell), which has been reconsidered with the emergence of composites structures~\cite{Heyliger92, Chen01, Ye14}. Besides because the eigenmodes are intrinsic to the structure they can be used to characterize unknown materials or geometries, using \textit{e.g.} the resonant scattering theory for immersed spheres \cite{Williams85} or Raman scattering for nanoparticles \cite{Saviot04}. Sphere eigenmodes have also found applications in the non-destructive testing of ceramic balls in aeronautics \cite{Petit05}.

Eigenmodes are the solutions of a dispersion relationship, which can be obtained analytically in a homogeneous sphere. However advanced numerical methods are required when the structure is complex (\textit{e.g.} multilayered). \citet{Buchanan02} have proposed a simple model based on two-dimensional finite elements. However the computational cost raises quickly with the frequency of interest, and this method is therefore rather limited to low-frequency computation. On the contrary, semi-analytical methods yield a one-dimensional model, and thus the cost remains reasonable even at high frequencies. The basic idea of semi-analytical methods is to describe analytically some directions (in a sphere, the angular ones) and to discretize the remaining one (the radial direction). 

This principle has been applied with spherical harmonics along the angular directions and finite elements along the radial direction by \citet{Heyliger92} and \citet{Park02}. It leads to a linear eigenvalue problem which is simple to solve. However, both existing formulations do not take full advantage of the analytical description of the solution along the angular coordinates. The eigenproblem of \citet{Heyliger92} is not given in a closed-form, so that numerous integrations must be performed before computing the modes. The model of \citet{Park02} (the so-called spherical thin layer method (STLM)) is obtained for linear and quadratic radial interpolation only. It cannot be readily extended to other interpolating functions and discretization techniques, such as spectral elements for instance. It is then of interest to propose a more general formulation.

From the authors' point of view, the main issue is to properly identify the orthogonality relationships of spherical harmonics which must be used to preserve the separation of radial and angular variables in the elastodynamic balance equations. These equilibrium equations correspond to vector wave equations, which complicates the problem. As will be outlined in this paper, two kinds of orthogonality relationships are necessary to eventually obtain a general semi-analytical formulation. On one hand, the orthogonality of vector spherical harmonics is needed for the kinetic energy term (including the scalar product of displacements). This first kind of orthogonality is rather well-known in the literature \cite{Kausel06}. On the other hand, the orthogonality of tensor spherical harmonics is required for the potential energy term (including the stress-strain tensor product). This second kind of orthogonality is much more mathematically involved (see \textit{e.g.} Refs. \cite{James76, Martinec00}).

As far as wave propagation is concerned, surface acoustic waves on a sphere are strongly related to the eigenmodes. The latter corresponds to standing waves which naturally occur because of the closed geometry of the sphere. Any wave can be reconstructed by a superposition on the eigenmodes \citep{Aki80, Eringen75a}. The Rayleigh surface wave, when excited by a point source, propagates all over the sphere (because of diffraction) and merges at the pole opposite to the source (because of the curvature). However as shown by \citet{Tsukahara00}, \citet{Ishikawa03} and \citet{Clorennec04}, if the source is a line of specific width, the Rayleigh wave is naturally collimated, that is, the wave is not diffracted but propagates with a quasi-constant width. Such a wave makes several roundtrips, which has been exploited to design gas sensors~\citep{Yamanaka09}. This phenomenon can be explained as a balance between diffraction and curvature effects. In this paper, we are interested in modelling the propagation of waves excited by an arbitrary source, based on the eigenvibrations of a multilayered sphere. In particular, we want to reproduce the collimation of a Rayleigh wave generated by a line source in a homogeneous sphere, and to investigate the perturbation induced by the addition of a thin viscoelastic coating to the sphere. 

The main objective of this paper is to elaborate a general one-dimensional semi-analytical finite element model to compute both the free (eigenmodes) and forced responses of an elastic sphere of complex internal structure (\textit{e.g.} multilayered). This model is presented in Sec.~\ref{sec:numerical_model} of this paper. The forced response model is calculated explicitly based on modal superposition on the eigenmodes. It is subsequently used to reconstruct the propagation of surface waves. In Sec.~\ref{sec:validation_section}, the computed eigenmodes are compared with literature results. The forced response is used in Sec.~\ref{sec:SAW} to simulate a collimating Rayleigh wave. The behaviour of this wave is analysed in terms of the eigenvibrations of the sphere. Finally, the effect of a viscoelastic coating on the collimating wave is investigated.

\section{The numerical model}
\label{sec:numerical_model}

\subsection{Elastodynamic variational formulation}

\begin{figure}
\centering
\includegraphics[width=0.30\textwidth, keepaspectratio=true]{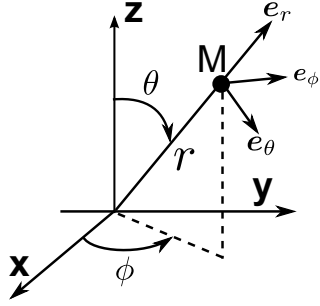}
\caption{Spherical coordinate system. $r$: radial coordinate; $0\leq\theta\leq\pi$: polar or colatitude angle; $0\leq\phi\leq2\pi$: azimuthal angle.}
\label{fig:spherical_frame}
\end{figure}

We consider a solid sphere of radius $r=a$. The problem is described in the spherical coordinate system $(r, \theta, \phi)$ shown in Fig.~\ref{fig:spherical_frame}. $r$ is the radial direction; $\theta$ is the polar or colatitude angle, with $0\leq\theta\leq\pi$; $\phi$ is the azimuthal angle, with $0\leq\phi\leq2\pi$. A time-harmonic dependance $\mathrm{e}^{-\mathrm{j}\omega t}$ is chosen for the displacement field, with $\vt{u}(r,\theta,\phi)=[u_r(r, \theta, \phi), \ u_\theta(r, \theta, \phi), \ u_\phi(r, \theta, \phi)]^{\mathrm{T}}$. The superscript $\mathrm{T}$ denotes matrix transpose. The elastodynamics variational formulation is (see \textit{e.g.} \citet[Chap. 4]{Bathe95}):
\begin{equation}
\int_{V} \delta\vt{\epsilon}^{\mathrm{T}}\vt{\sigma}\mathrm{d}V
-\omega^2\int_{V}\rho\delta\vt{u}^{\mathrm{T}}\vt{u}\mathrm{d}V 
=\int_{V}\delta\vt{u}^{\mathrm{T}}\vt{f}\mathrm{d}V+\int_{\partial V}\delta\vt{u}^{\mathrm{T}}\vt{t}\mathrm{d}\partial V\,,
\label{eq:variational_formulation}
\end{equation}
with $\mathrm{d}V=r^2\mathrm{d}r\sin\theta\mathrm{d}\theta\mathrm{d}\phi$. $\vt{f}$ is the vector of volumic acoustic forces. $\vt{t}$ is the vector of stresses on a spherical isosurface of radius $r=a$, such that $\mathrm{d}\partial V=r^2\sin\theta\mathrm{d}\theta\mathrm{d}\phi$ (nota that a stress boundary condition may also be applied on the inner radius, supposed at $r=b$, in case of a hollow sphere). The stress and strain vectors are respectively given by $\vt{\sigma}=[\sigma_{rr}, \ \sigma_{\theta\theta},\ \sigma_{\phi\phi},\ \sigma_{\theta\phi},\ \sigma_{r\phi},\ \sigma_{r\theta}]^{\mathrm{T}}$ and $\vt{\epsilon}=[\epsilon_{rr}, \ \epsilon_{\theta\theta},\epsilon_{\phi\phi},\ 2\epsilon_{\theta\phi},\ 2\epsilon_{r\phi},\ 2\epsilon_{r\theta}]^{\mathrm{T}}$. The stress-strain relation is $\vt{\sigma}=\vt{C}\vt{\epsilon}$. $\vt{C}$ is the matrix of material properties. The materials can be elastic or viscoelastic. The model is restricted to the case of transversely isotropic materials, such that:
\begin{equation}
\vt{C}=\begin{bmatrix}
 C_{11} & C_{12} & C_{12} & 0 & 0 & 0\\
  C_{12} & C_\alpha & C_{23} & 0 & 0 & 0\\ 
  C_{12} & C_{23} & C_\alpha & 0 & 0 & 0\\ 
  0 & 0 & 0 & C_{44} & 0 & 0\\ 0 & 0 & 0 & 0 & C_{55} & 0\\ 0 & 0 & 0 & 0 & 0 & C_{55}
\end{bmatrix}\,,
\label{eq:C_matrix}
\end{equation}
where $C_\alpha=2 C_{44} + C_{23}$. This restriction is necessary to allow the separation of angular and radial variables in Sec.~\ref{ssec:vectorSHT}. Besides, material properties can vary arbitrarily along the radius only (\textit{e.g.} a sphere made of several homogeneous layers).

The strain-displacement relation is $\vt{\epsilon}=\vt{L}\vt{u}$. The operator $\vt{L}$ is given by:
\begin{equation}
\vt{L}=\vt{L}_r\frac{\partial}{\partial r} +\vt{L}_\theta \frac{1}{r}\frac{\partial}{\partial \theta}+\vt{L}_\phi\frac{1}{r\sin\theta}\frac{\partial}{\partial \phi}+\frac{1}{r}\vt{L}_{1}+\frac{\cot \theta}{r}\vt{L}_{2}\,,
\end{equation}
with:
\begin{equation}
\vt{L}_r=\begin{bmatrix}
1 & 0 & 0 \\
0 & 0& 0 \\
0 & 0& 0 \\
0& 0& 0\\
0 &0 & 1 \\
0 &1& 0
\end{bmatrix}\,, \, 
\vt{L}_\theta = \begin{bmatrix}
0 & 0 & 0\\
0 & 1& 0 \\
0 &0& 0\\
0 & 0& 1\\
0 & 0& 0 \\
1& 0& 0
\end{bmatrix}\,, \, 
\vt{L}_\phi =\begin{bmatrix}
0 & 0& 0\\
0& 0& 0\\
0 &0& 1\\
0 & 1& 0\\
1& 0& 0\\
0 & 0& 0
\end{bmatrix}\,,
\vt{L}_{1}=\begin{bmatrix}
0&0&0\\
1&0&0\\
1&0&0\\
0&0&0\\
0&0&-1\\
0&-1&0
\end{bmatrix}\,, \,
\vt{L}_{2}=\begin{bmatrix}
0&0&0\\
0&0&0\\
0&1&0\\
0&0&-1\\
0&0&0\\
0&0&0
\end{bmatrix}\,.
\label{eq:L_operator}
\end{equation}

\subsection{Vector spherical harmonic expansion}
\label{ssec:vectorSHT}

Let us first remind that, applying the Helmholtz decomposition theorem, finding the solutions of the homogeneous elastodynamic equations (strong formulation) consists in solving three scalars Helmholtz equations. The radial and angular coordinates can be separated, and the angular scalar dependence is written on a basis of spherical harmonic functions~\citep{Eringen75a, Aki80, Kausel06}. The normalized scalar spherical harmonic functions are given by:
\begin{equation}
Y_l^m(\theta,\phi)=\frac{N_l^m}{\sqrt{2 \pi}}P_l^m(\cos\theta)\mathrm{e}^{\mathrm{j}m\phi}\,, \label{eq:norm_spherical_harmonic}
\end{equation}
with the degree $l$ $(l\geq 0)$ and the order $m$ $(|m|\leq l)$~\citep{Arfken99}. $N_l^m=\sqrt{\frac{(2l+1)(l-m)!}{2(l+m)!}}$ is the normalization factor. The integers $l$ and $m$ are also called the polar and azimuthal wavenumbers~\citep{Aki80}. $P_l^m(\cos\theta)$ is the associated Legendre polynomial of the first kind, which satisfies the Legendre equation~\citep{Arfken99}:
\begin{equation}
\frac{\mathrm{d}^2 P_l^m}{\mathrm{d}\theta^2}+\cot\theta\frac{\mathrm{d}P_l^m}{\mathrm{d}\theta}+\left(\overline{l}-\frac{m^2}{\sin^2\theta}\right)P_l^m=0\,,
\label{eq:legendre_equation}
\end{equation} 
with $\overline{l}=l(l+1)$. The explicit form of associated Legendre polynomials is written by convention~\citep{Arfken99}:
\begin{equation}
P_l^m(\cos\theta)=(-1)^m(\sin\theta)^m\frac{\mathrm{d}^m}{\mathrm{d}(\cos\theta)^m}P_l(\cos\theta)\,,
\end{equation}
where $P_l(\cos\theta)$ is the Legendre polynomial of the first kind, including the Condon-Shortley phase $(-1)^m$.

From the scalar solutions, one can eventually obtain the vector displacement solution in which radial and angular variables are also separated. In this paper, the radial behaviour of the displacement is interpolated using finite elements (see Sec.~\ref{ssec:FE_interpolation}) instead of using exact spherical Bessel function, while the analytical description of the angular behaviour, using vector spherical harmonics, is exploited.

In Eq.~\eqref{eq:variational_formulation}, the displacement field is then written as:
\begin{equation}
\vt{u}(r,\theta,\phi)=\sum_{l \geq 0} \sum_{|m|\leq l}\vt{S}_l^m(\theta,\phi)\hat{\vt{u}}_l^m(r)\,,
\label{eq:SH_decomposition_u}
\end{equation}
where $\hat{\vt{u}}_l^m(r)=[\hat{u}_l^m(r),\ \hat{v}_l^m(r),\ \hat{w}_l^m(r)]^{\mathrm{T}}$ is the vector of the $(l,m)$--coefficients of the expansion (to be determined). The matrix $\vt{S}_l^m$ conveniently concatenates the vector spherical harmonics and is given by \cite{Kausel06,Park02}:
\begin{equation}
\vt{S}_l^m(\theta,\phi) = \begin{bmatrix}
\mathrm{Y}_l^m(\theta,\phi) & 0 & 0\\ 
0 & \frac{\partial \mathrm{Y}_{l}^m(\theta,\phi)}{\partial \theta}& -\frac{\partial \mathrm{Y}_{l}^m(\theta,\phi)}{\sin \theta\partial\phi}\\ 
0 & \frac{\partial \mathrm{Y}_{l}^m(\theta,\phi)}{\sin \theta\partial\phi} & \frac{\partial \mathrm{Y}_{l}^m(\theta,\phi)}{\partial \theta}
\end{bmatrix}\,.
\label{eq:Slm}
\end{equation}
The same form is also assumed for the volumic forces and the normal stresses, that is:
\begin{align}
\vt{f}(r,\theta,\phi)&=\sum_{l \geq 0} \sum_{|m|\leq l}\vt{S}_l^m(\theta,\phi)\hat{\vt{f}}_l^m(r)\,,\label{eq:SH_decomposition_f} \\ \vt{t}(\theta,\phi)&=\sum_{l \geq 0} \sum_{|m|\leq l}\vt{S}_l^m(\theta,\phi)\hat{\vt{t}}_l^m\,.
\label{eq:SH_decomposition_t}
\end{align} 

It is noteworthy that the vector spherical harmonics form an orthogonal basis~\citep{Martinec00,Kausel06}, with:
\begin{equation}
\int_0^\pi \int_0^{2\pi}\vt{S}_k^{p*}\vt{S}_l^m\mathrm{d}\phi\sin\theta\mathrm{d}\theta=\begin{bmatrix}
1& 0 & 0\\
0 & \overline{l} & 0 \\
0 & 0 & \overline{l}
\end{bmatrix}\delta_{kl}\delta_{mp}\,,
\label{eq:orth_Slm}
\end{equation}
where $*$ stands for the transpose conjugate matrix. The proof is briefly recalled in \ref{app:orthogonality}. As explained later, the key point of the formulation proposed in this paper is then to choose the test function as follows:
\begin{equation}
\delta\vt{u}^{\mathrm{T}}(r,\theta,\phi)=\delta\hat{\vt{u}}^{\mathrm{T}}(r)\vt{S}_k^{p*}(\theta,\phi)\,.
\label{eq:FE_form_test_function}
\end{equation}
From Eq.~\eqref{eq:FE_form_test_function}, one can write the virtual strains as: $\delta\vt{\epsilon}^{\mathrm{T}}=\left[\vt{L}\vt{S}_k^{p*\mathrm{T}}\delta\hat{\vt{u}}\right]^{\mathrm{T}}$.

\subsection{Finite element approximation of the radial dependance of wavefields}
\label{ssec:FE_interpolation}

Along the radial direction, a finite element approximation is applied such that the displacement on each element is given by:
\begin{equation}
\hat{\vt{u}}_l^{m,e}(r)=\vt{N}^e(r)\hat{\vt{U}}_l^{m,e}\,.
\label{eq:FE_discretization}
\end{equation}
$\vt{N}^e(r)$ is the matrix of one-dimensional interpolation functions. $\hat{\vt{U}}_l^{m,e}$ is the vector of nodal displacements.

In Eq.~\eqref{eq:variational_formulation}, the angular integrations are then computed and simplified using Eq.~\eqref{eq:orth_Slm}. To evaluate the integral $\int \delta\vt{\epsilon}^{\mathrm{T}}\vt{\sigma}\sin\theta\mathrm{d}\theta\mathrm{d}\phi$, additional relationships coming from the orthogonality of tensor spherical harmonics are necessary. These are given in \ref{app:orthogonality}. A detailed example is also given in \ref{app:example} for the calculation of one matrix component. Finally, after lengthy algebraic manipulations, the following global matrix system can be obtained:
\begin{equation}
\left(\vt{K}(l)-\omega^2\vt{M}(l)\right)\hat{\vt{U}}_l^m=\hat{\vt{F}}_l^m\,.
\label{eq:FE_system}
\end{equation}
The stiffness matrix is given by:
\begin{equation}
\vt{K}(l)=\vt{K}_{1}(l)+\vt{K}_{2}(l)+\vt{K}_{2}^{\mathrm{T}}(l)+\vt{K}_{3}(l)\,,
\end{equation}
where elementary matrices are:
\begin{align}
\vt{K}_{1}^e(l)&= \int \frac{\mathrm{d}\vt{N}^{e\mathrm{T}}}{\mathrm{d}r}
\begin{bmatrix}
C_{11}& 0 &0 \\
0 & \overline{l}C_{55} & 0 \\
0 &0& \overline{l}C_{55} 
\end{bmatrix} \frac{\mathrm{d}\vt{N}^e}{\mathrm{d}r}r^2\mathrm{d}r
\,,\label{eq:elementary_matrixK1}\\
  \vt{K}_{2}^e(l)&=\int \frac{\mathrm{d}\vt{N}^{e\mathrm{T}}}{\mathrm{d}r}\begin{bmatrix}
2C_{12}& -\overline{l}C_{12} &0 \\
\overline{l}C_{55} & -\overline{l}C_{55} & 0 \\
0 &0& -\overline{l}C_{55} 
\end{bmatrix}\vt{N}^er\mathrm{d}r\,,\label{eq:elementary_matrixK2} \\
\vt{K}_{3}^e(l)&=\int \vt{N}^{e\mathrm{T}}
\left[\begin{matrix}
\overline{l}C_{55}+4C_\beta &  -\overline{l}\big(C_{55}+2C_\beta\big) & 0\\
-\overline{l}\big(C_{55}+2C_\beta\big)&\overline{l}(C_{55}+\overline{l}C_{23}+2(\overline{l}-1)C_{44})& 0\\
0 & \overline{l}(C_{55}+(\overline{l}-2)C_{44}) & 0\end{matrix}\right]\vt{N}^e\mathrm{d}r\,,\label{eq:elementary_matrixK3}
\end{align} 
where $C_\beta=C_{23}+C_{44}$.

The elementary mass matrix is given by:
\begin{equation}
\vt{M}^e(l)=\int\rho\vt{N}^{e\mathrm{T}}\begin{bmatrix}
1 & 0 & 0 \\ 0& \overline{l}& 0 \\ 0& 0 & \overline{l}
\end{bmatrix}\vt{N}^er^2\mathrm{d}r\,.
\label{eq:mass_matrix}
\end{equation}
The force vector $\hat{\vt{F}}_l^m$ gathers the contribution of volumic forces and stresses, that is:
\begin{equation}
\hat{\vt{F}}_l^m = \hat{\vt{F}}_{l,v}^m+\hat{\vt{F}}_{l,s}^m\,.
\label{eq:FE_source}
\end{equation}
The elementary volumic forces are given by:
\begin{equation}
 \hat{\vt{F}}_{l,v}^{m,e}=\int\vt{N}^{e\mathrm{T}}\begin{bmatrix}
1 & 0 & 0 \\ 0& \overline{l}& 0 \\ 0& 0 & \overline{l}\end{bmatrix}\hat{\vt{f}}_l^{m,e}r^2\mathrm{d}r\,.
\end{equation} 
The contribution of stresses can be written:
\begin{equation}
\delta\hat{\vt{U}}^{\mathrm{T}}\hat{\vt{F}}_{l,s}^m=\begin{bmatrix} 1 & 0 & 0 \\ 0& \overline{l}& 0 \\ 0& 0 & \overline{l}
\end{bmatrix}
\left[\delta\hat{\vt{u}}r^2\hat{\vt{t}}_l^m\right]_{r=b}^{r=a}\,,
\end{equation}
where $\delta\hat{\vt{U}}$ is the vector of virtual nodal displacements. 

As a final remark, let us highlight the significance of the test function \eqref{eq:FE_form_test_function}. Owing to this choice, the orthogonality of vector spherical harmonics (Eq.~\eqref{eq:orth_Slm}) immediately appears in each integral of Eq.~\eqref{eq:variational_formulation}. Moreover, this choice also takes advantage of the orthogonality of tensor spherical harmonics \cite{James76, Martinec00}, which leads to the identities summed up by Eqs.~\eqref{eq:ortho_tensor1}--\eqref{eq:ortho_tensor2} in \ref{app:orthogonality}. Both vector and tensor orthogonality relationships are mandatory to get uncoupled governing equations for each pair of wavenumbers $(l,m)$, as eventually obtained in Eq.~\eqref{eq:FE_system}. This yields a general formulation with a fully analytical description of the problem along the two angular coordinates. This is not the case in the model of \citet{Heyliger92}, in which angular integrals must be solved analytically or numerically for each value of $l$. On the other hand, compared with the formulation of \citet{Park02}, the expressions of the matrices given by Eqs.~\eqref{eq:elementary_matrixK1}--\eqref{eq:elementary_matrixK3} are valid for any choice of interpolation functions.

\subsection{The source-free problem: computation of the eigenmodes}
To compute the eigenmodes of the sphere, the source-free problem must be considered (\textit{i.e.} $\hat{\vt{F}}_l^m=\vt{0}$ in Eq.~\eqref{eq:FE_system}). It yields a standard linear eigenproblem in terms of $\omega$. For each integer value of $l$, one obtains $N$ eigenfrequencies $\omega_l^{(n)}$ and eigenvectors $\hat{\vt{U}}_l^{(n)}$ (corresponding to the radial mode shapes), with $n=1 \ldots N$. 

Let us stress than when this problem is solved analytically, the eigenfrequencies are the roots of transcendental equations involving spherical Bessel functions (the dispersion relationship can be found for a homogeneous and isotropic sphere in Ref. \cite{Eringen75a}), which must be solved with root-finding algorithms. Their convergence can be poor at high frequencies because of instabilities~\citep{Fong05}. Conversely, the linear eigenproblem obtained in this paper can be solved with standard methods.

The eigenproblem shares some properties with the analytical dispersion relationship. Both are degenerate with respect to the azimuthal wavenumber $m$, such that there is $2l+1$ independent modes with the same eigenfrequency $\omega_l^{(n)}$~\citep{Silbiger62,Duffey07}. Furthermore, it can be noticed from the structures of matrices in Eqs.~\eqref{eq:elementary_matrixK1}--\eqref{eq:mass_matrix} that the eigensystem can be readily divided into two independent linear eigenproblems, namely:
\begin{align}
(\vt{K}_S-\omega^2\vt{M}_S)\hat{\vt{U}}_S &= \vt{0}\,,\label{eq:EVP_spheroidal}  \\
(\vt{K}_T-\omega^2\vt{M}_T)\hat{\vt{U}}_T & = \vt{0}\,.\label{eq:EVP_toroidal}
\end{align}
One recovers the two families of eigenmodes in a sphere~\citep{Eringen75a}. The first eigenproblem yields the so-called spheroidal modes, which are polarized in every direction. The second eigenproblem yields the so-called torsional modes, which are polarized only in the angular directions. For the simplicity of the formulation, this decomposition is not exploited in this paper (the finite element discretization is only one-dimensional and leads to fast computations).

\subsection{The forced response: wave propagation}
\label{ssec:forced_response}

To reconstruct surface wave propagation, the forced response problem must be considered. First, let us write the linear eigenproblem for a given mode $(l,n)$:
\begin{equation}
(\vt{K}(l)-\omega_l^{(n)2}\vt{M}(l))\hat{\vt{U}}_l^{(n)} = \vt{0}\,.
\label{eq:eigensystem}
\end{equation}
Owing to viscoelasticity (if any), the matrices $\vt{K}$ and $\vt{M}$ are complex-valued and not Hermitian. However, both matrices are symmetric such that the following orthogonality relationships hold:
\begin{align}
\hat{\vt{U}}_l^{(k)\mathrm{T}}\vt{K}(l)\hat{\vt{U}}_l^{(n)}&=\omega_l^{(n)2}\delta_{kn}\,, \label{eq:EVP_ortho1}\\
\hat{\vt{U}}_l^{(k)\mathrm{T}}\vt{M}(l)\hat{\vt{U}}_l^{(n)}&=\delta_{kn}\,.\label{eq:EVP_ortho2}
\end{align}
Introducing the modal expansion $\hat{\vt{U}}_l^m=\sum_{n=1}^N \alpha^{(n)}\hat{\vt{U}}_l^{(n)}$ into the forced response equation \eqref{eq:FE_system}, multiplying by $\hat{\vt{U}}_l^{(k)\mathrm{T}}$ and using the orthogonality relationships \eqref{eq:EVP_ortho1} and \eqref{eq:EVP_ortho2} yields:
\begin{equation}
\alpha^{(n)}=\frac{\hat{\vt{U}}_l^{(n)\mathrm{T}}\hat{\vt{F}}_l^m}{\omega_l^{(n)2}-\omega^2}\,,
\end{equation}
such that:
\begin{equation}
\hat{\vt{U}}_l^m=\sum_{n=1}^N\frac{\hat{\vt{U}}_l^{(n)\mathrm{T}}\hat{\vt{F}}_l^m\hat{\vt{U}}_l^{(n)}}{\omega_l^{(n)2}-\omega^2}\,.
\label{eq:FRF}
\end{equation}
This quantity corresponds at each node to the displacement Frequency Response Function (FRF) of a pair $(l,m)$. Its inverse Fourier transform yields the transient displacement:
\begin{equation}
\hat{\vt{U}}_l^m(t) = \frac{1}{2\pi}\int_{-\infty}^{+\infty} \left[\sum_{n=1}^N\frac{\hat{\vt{U}}_l^{(n)\mathrm{T}}\hat{\vt{F}}_l^m(\omega)\hat{\vt{U}}_l^{(n)}}{\omega_l^{(n)2}-\omega^2}\right]\mathrm{e}^{-\mathrm{j}\omega t}\mathrm{d}\omega\,.
\label{eq:transient_l^m}
\end{equation}
Finally, the vector of physical nodal displacements in the time domain is given by:
\begin{equation}
\vt{U}(\theta,\phi,t)=\sum_{l \geq 0} \sum_{|m|\leq l} \vt{S}_l^m(\theta,\phi)\hat{\vt{U}}_l^m(t)\,.
\label{eq:transient_U}
\end{equation}

\subsection{Wave properties}
Phase and group velocities can be derived from the eigenfrequencies. The phase velocity is given by:
\begin{equation}
v_{p_l}^{(n)} =\frac{\text{Re}(\omega_l^{(n)}) a}{l+\frac{1}{2}}\,,
\end{equation}
where $ka=l+1/2$ accounts for the polar phase-shift of surface waves~\citep{Brune61, Eringen75a}.

The group velocity is defined as~\citep{Eringen75a}:
\begin{equation}
v_{g_l}^{(n)}=\text{Re} \frac{\partial \omega_l^{(n)}}{\partial l}\,.
\end{equation}
Following \citet{Finnveden04a}, the group velocity can be obtained from the finite element matrices (which avoids complex mode sorting). Deriving Eq.~\eqref{eq:eigensystem} with respect to $l$ yields:
\begin{equation}
\begin{split}
&\left(\frac{\partial \vt{K}(l)}{\partial l}-2\omega_l^{(n)}\vt{M}(l)\frac{\partial \omega_l^{(n)}}{\partial l}-\omega_l^{(n)2}\frac{\partial \vt{M}(l)}{\partial l}\right)\hat{\vt{U}}_l^{(n)}
\\&+(\vt{K}(l)-\omega_l^{(n)2}\vt{M}(l))\frac{\partial \hat{\vt{U}}_l^{(n)}}{\partial l}=0\,.
\end{split}
\end{equation}
Multiplying by $\hat{\vt{U}}_l^{(n)\mathrm{T}}$, the second term is equal to zero. One readily obtains the following expression:
\begin{equation}
v_{g_l}^{(n)}=\text{Re} \left[ \frac{\hat{\vt{U}}_l^{(n)\mathrm{T}}\left(\frac{\partial \vt{K}(l)}{\partial l}-\omega_l^{(n)2}\frac{\partial \vt{M}(l)}{\partial l} \right) \hat{\vt{U}}_l^{(n)}}{2\omega_l^{(n)}\hat{\vt{U}}_l^{(n)\mathrm{T}}\vt{M}(l)\hat{\vt{U}}_l^{(n)}} \right]\,.
\label{eq:vgroup}
\end{equation}
To compute the derivative of the mass and the stiffness matrices, it is noteworthy that they can be readily factored as:
\begin{align}
\vt{K}(l)& =\vt{K}_{1}^{'}+\overline{l}(\vt{K}_{2}^{'}+\vt{K}_{2}^{'\mathrm{T}})+\overline{l}^2\vt{K}^{'}_{3}\,, \\
\vt{M}(l)& = \vt{M}^{'}_{1}+\overline{l}\vt{M}^{'}_{2}\,.
\end{align}
Therefore, their derivatives are given by:
\begin{align}
\frac{\partial \vt{K}(l)}{\partial l}& =(2l+1)(\vt{K}^{'}_{2}+\vt{K}_{2}^{'\mathrm{T}})+2\overline{l}(2l+1)\vt{K}^{'}_{3}\,, \\
\frac{\partial \vt{M}(l)}{\partial l}& = (2l+1)\vt{M}^{'}_{2}\,.
\end{align}

\subsection{Remarks on the sherical harmonic expansion}
\label{ssec:rq_SHT}

In Eqs.~\eqref{eq:SH_decomposition_u}, \eqref{eq:SH_decomposition_f}, \eqref{eq:SH_decomposition_t}, the quantities $\hat{\vt{u}}_l^m,\ \hat{\vt{f}}_l^m,\ \hat{\vt{t}}_l^m$ stand for the coefficients of a Vector Spherical Harmonic Transform (Vector SHT analysis), given by:
\begin{equation}
\hat{\vt{u}}_l^m(r)=\int_0^\pi \int_0^{2\pi}\vt{S}_l^{m*}(\theta,\phi)\vt{u}(r,\theta,\phi)\mathrm{d}\phi\sin\theta\mathrm{d}\theta\,.
\end{equation}
Accordingly the physical quantities $\vt{u}_l^m,\ \vt{f}_l^m,\ \vt{t}_l^m$ are the results of an Inverse Vector Spherical Harmonic Transform (Vector SHT synthesis). As shown by \citet{Kostelec00}, the $\theta$-derivative of $Y_l^m$ in $\vt{S}_l^m$ can be related recursively to the $l+1$ and $l-1$ spherical harmonics degrees, such that the vector SHT is equivalent to several scalar SHT. For any scalar function $h(\theta,\phi)$, its SHT synthesis is:
\begin{equation}
h(\theta,\phi)=\sum_{l \geq 0} \sum_{|m|\leq l}Y_l^m(\theta,\phi)\hat{h}_l^m\,.
\label{eq:def_synthesis}
\end{equation}
The complex-valued coefficients $\hat{h}_l^m$ can be obtained from the SHT analysis of the function $h$, that is:
\begin{equation}
\hat{h}_l^m =\int_0^\pi \int_0^{2\pi}Y_l^{m*}(\theta,\phi)h(\theta,\phi)\mathrm{d}\phi\sin\theta\mathrm{d}\theta\,.
\label{eq:def_analysis}
\end{equation}
These transforms cannot be evaluated analytically in general. Several accurate and quick numerical tools have been proposed in the literature~\citep{Healy03,Wieczorek18,Schaeffer13}. In this paper, the SHT analysis and synthesis are performed following the numerical strategy described in Refs. \cite{Wieczorek18} and \cite{Schaeffer13}. The fundamental steps are briefly recalled in the following for self-consistency.

First, Eq.~\eqref{eq:def_analysis} can be written as:
\begin{equation}
\hat{h}_l^m =\frac{N_l^m}{\sqrt{2\pi}}\int_0^\pi \left[\int_0^{2\pi}\mathrm{e}^{-\mathrm{j}m\phi}f(\theta,\phi)\mathrm{d}\phi\right]P_l^m(\cos\theta)\sin\theta\mathrm{d}\theta\,.
\label{eq:decomposition_analysis}
\end{equation}
Equation \eqref{eq:decomposition_analysis} shows that the SHT analysis can be subdivided into a Fourier transform followed by a projection on the basis of associated Legendre polynomials of the first kind. 

The Fourier transform integral is computed using a Discrete Fourier Transform (DFT) on a minimum of $N_{TF}=2L+1$ samples along the azimuthal coordinate according to Shannon's theorem, where $L$ is the maximum value of $l$. In practice, the DFT can be efficiently computed using Fast Fourier Transform (FFT) algorithms~\citep{Frigo98}. It yields the spectrum of coefficients $h^m(\theta)$.

The projection on the basis of associated Legendre polynomial is then evaluated using a Gauss-Legendre quadrature (GLQ). For a given $m$ and using the change of variable $x=\cos \theta$, one obtains:
\begin{equation}
 \int_0^\pi h^m(\theta)P_l^m(\cos\theta)\sin\theta\mathrm{d}\theta =\sum_{q=1}^{L+1} w_qh^m(\acos x_q)P_l^m(x_q)\,, \label{eq:GLQ}
\end{equation}
where $w_q$ are the Gauss weights and $x_q$ are the Gauss points. Using $L+1$ Gauss points, the integration is exact if the product $h^m(\acos x_q)P_l^m(x_q)$ is a polynomial of maximum degree $2L$. The latter assumption is not strictly verified, but the accuracy has been shown to be very good in practice~\cite{Wieczorek18}.

The SHT synthesis \eqref{eq:def_synthesis} can be written as:
\begin{equation}
h(\theta,\phi)=\frac{N_l^m}{\sqrt{2\pi}}\sum_{m=-L}^{m=L} \left[ \sum_{l = |m|}^L \hat{h}_l^mP_l^m(\cos\theta)\right] \mathrm{e}^{\mathrm{j}m\phi}\,.
\label{eq:switch_synthesis}
\end{equation}
This equation shows that for each value of $\theta$ the first step of the SHT synthesis is a summation over the associated Legendre polynomial basis, followed by an Inverse DFT. Note that the number of synthesis along $\theta$ can be reduced taking advantage of the symmetry of the associated Legendre polynomial across the equator, with the identity:
\begin{equation}
 P_l^m(\cos(\pi-\theta))=(-1)^{(l+m)}P_l^m(\cos\theta)\,.
\end{equation}

\section{Validation test case: free vibrations of an isotropic homogeneous sphere}
\label{sec:validation_section}

\begin{table*}[ht]
\makebox[\textwidth][c]
{
\begin{tabular}{ccccccc}
\hline
\hline 
Material & $E$ (\giga\pascal) & $\rho$ (\kilo\gram\usk\rpcubic\meter) & $c_l$ (\meter\usk\reciprocal\second) & $c_s$ (\meter\usk\reciprocal\second) & $\eta_l$ (Np\usk\reciprocal{wavelength}) &  $\eta_s$ (Np\usk\reciprocal{wavelength}) \\ 
\hline
Steel & 200 & 7932 & 5500.7 & 3175.8 & 0.003 & 0.008 \\
Epoxy &  9 & 1600 & 2960 & 1450 & 0.0047 & 0.0069 \\
\hline 
\hline
\end{tabular} 
}
\caption{Material properties}
\label{tab:materials}
\end{table*}

\subsection{Description of the test case}

Let us consider a surface-free isotropic sphere of radius $a=\unit{10}{\milli\meter}$. The sphere is made of steel. Materials properties are given in Table \ref{tab:materials} (here the viscoelastic parameters are equal to zero). The material is isotropic and its stiffness matrix can be written as:
 \begin{equation}
\vt{C}=\begin{bmatrix}
 3\lambda & \lambda & \lambda & 0 & 0 &0\\
 \lambda & 3\lambda & \lambda & 0 & 0& 0\\
 \lambda & \lambda & 3\lambda & 0 & 0 &0\\
 0& 0& 0 & \lambda & 0 & 0 \\
 0 &0& 0& 0& \lambda & 0\\
 0& 0& 0& 0& 0& \lambda
\end{bmatrix}\,,
\end{equation}
with the Lamé's parameter $\lambda=E\nu/[(1+\nu)(1-2\nu)]$ ($\nu=0.25$). 

The eigenmodes computed with the numerical method of this paper are compared with the results of Ref. \cite[Ch. 8]{Eringen75a} for $n=1, \ldots, 5$ and $l=1,\ldots,60$. The non-dimensional eigenfrequencies are defined by $\overline{\omega}_l^{(n)}=a\omega_l^{(n)}/c_S$, where $c_S$ is the shear wave velocity. The radius is discretized with three-nodes line elements of length $\Delta r = 0.012a$, which corresponds to a sixth of the minimum radial wavelength given by $2\pi c_S/\text{max}(\omega_l^{(n)})$. The one-dimensional numerical model then comprises 1014 degrees of freedom (dofs). 
  
\subsection{Results}
\label{sec:validation_results}

\begin{figure*}
\centering
 \includegraphics[width=0.85\textwidth, keepaspectratio=true]{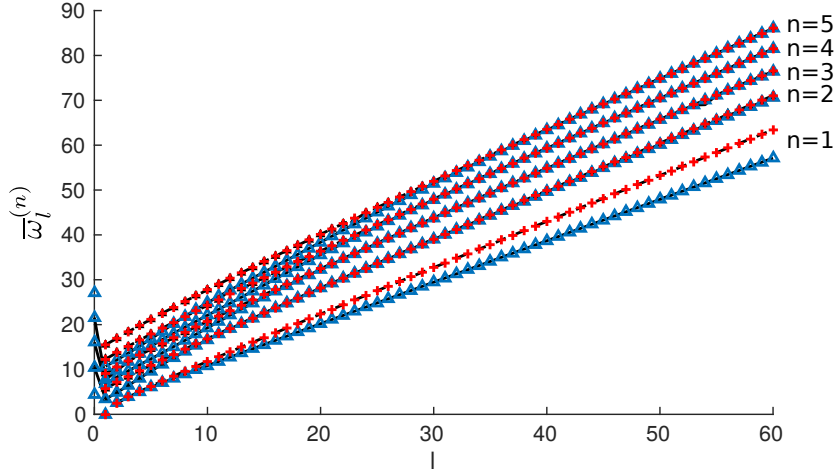}
 \caption{Non-dimensional eigenfrequencies $\overline{\omega}_l^{(n)}$ of a surface-free sphere made of steel. Solid and dashed blacked lines: results of \citet{Eringen75a} for spheroidal and torsional modes. Blue triangles: numerical results for spheroidal modes. Red crosses: numerical results for torsional modes.} 
 \label{fig:eigenfrequencies_validation}
\end{figure*}

Figure \ref{fig:eigenfrequencies_validation} compares the eigenfrequencies of Ref. \cite{Eringen75a} and those obtained with the numerical method of this paper. Both results are superimposed. The curves of the spheroidal modes (blue triangles in Fig. \ref{fig:eigenfrequencies_validation}) start at $l=0$ at breathing mode eigenfrequencies (modes with a radial polarization only). At low wavenumber $l$, these curves also exhibit a sudden change of slope, characterizing a strong dispersive behaviour.

\begin{figure*}
\centering
 \includegraphics[width=0.95\textwidth, keepaspectratio=true]{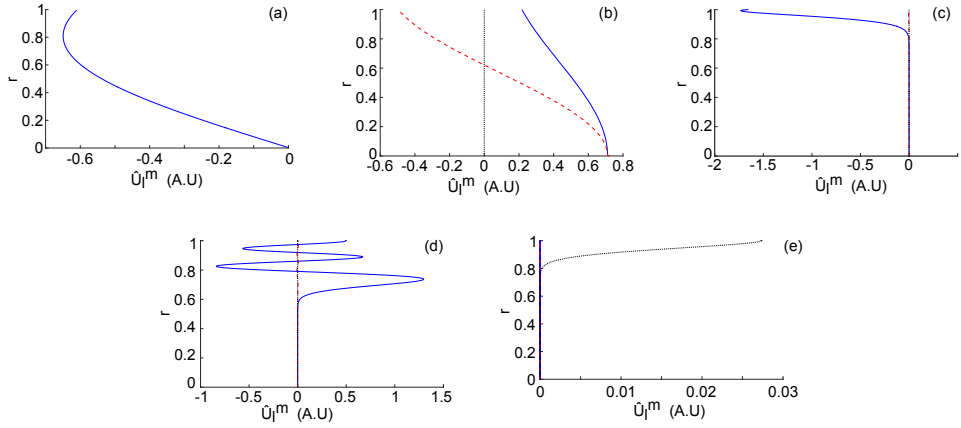}
 \caption{Radial modeshapes (arbitrary unit) $\hat{u}_l^m(r)$ (solid blue line), $\hat{v}_l^m(r)$ (dashed red line) and $\hat{w}_l^m(r)$ (dotted black line) of (a) spheroidal mode, $l=0$, $n=1$ ($\overline{\omega}_0^{(1)}= 4.44$); (b) spheroidal mode, $l=1$, $n=2$ ($\overline{\omega}_1^{(2)}= 3.412$); (c) spheroidal mode, $l=60$, $n=1$ ($\overline{\omega}_{60}^{(1)}= 57.13$); (d) spheroidal mode, $l=60$, $n=5$ ($\overline{\omega}_{60}^{(5)}= 86.03$); (e) torsional mode, $l=60$, $n=1$ ($\overline{\omega}_{60}^{(1)}= 63.44$).} 
 \label{fig:radial_modeshape}
\end{figure*}

Some radial modeshapes $\vt{U}_l^m$ are displayed in Fig.~\ref{fig:radial_modeshape}. The dofs $\hat{u}_l^m$, $\hat{v}_l^m$ and $\hat{w}_l^m$ can be related to physical displacements, based on the two independent eigensystems \eqref{eq:EVP_spheroidal}--\eqref{eq:EVP_toroidal}. The dofs $\hat{u}_l^m$ describe the radial dependence of $u_r$. As expected, they are null for torsional modes which are polarized along the angular directions only -- see Fig.~\ref{fig:radial_modeshape}e. The dofs $\hat{v}_l^m$ give the radial dependence of $u_\theta $ and $u_\phi$ for spheroidal modes. As shown in Fig.~\ref{fig:radial_modeshape}b,c,d, the dofs $\hat{w}_l^m$ are null in that case. On the contrary, the dofs $\hat{w}_l^m$ are the only non-zero dofs for torsional modes -- see Fig.~\ref{fig:radial_modeshape}e. To verify the accuracy on modeshapes, Fig.~\ref{fig:radial_modeshape}b can be compared with the results of Ref. \cite{Eringen75a}. A very good visual agreement is obtained.

The radial behaviour of the modes depends on their polar wavenumber $l$ and on their order $n$. For $l=0$, the motion is purely radial (breathing mode) and distributed over the radius (Fig.~\ref{fig:radial_modeshape}a). The spheroidal modes with $l=1$ are the only modes with a non-zero displacement at $r=0$ -- see Fig.~\ref{fig:radial_modeshape}b and \cite{Eringen75a}. As shown in Fig.~\ref{fig:radial_modeshape}c and \ref{fig:radial_modeshape}d, for high values of $l$ the displacement is confined near the surface, particularly for the fundamental mode with $n=1$ (Fig.~\ref{fig:radial_modeshape}c). When the order $n$ of the mode increases, \textit{e.g.} in Fig.~\ref{fig:radial_modeshape}d with $n=5$, the inner displacement increases and oscillations can be observed along the radius.

\begin{figure*}
\centering
\includegraphics[width=0.85\textwidth, keepaspectratio=true]{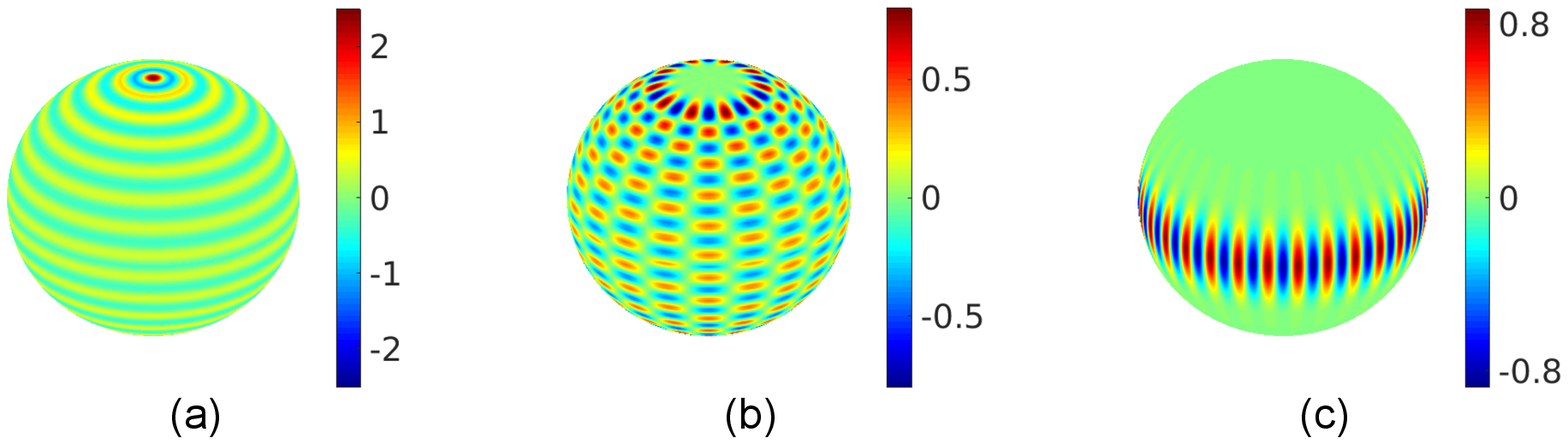}
\caption{Normal modal displacement (arbitrary unit) $u_r(r=a,\theta,\phi)$ of the fundamental spheroidal modes $l=30$, $n=1$ ($\overline{\omega}_{30}^{(1)}=29.46$) (a) for $m=0$; (b) for $m=10$; (c) for $m=l$.}
\label{fig:3D_modeshape}
\end{figure*}

A simple post-processing step enables to represent the modal displacement on a spherical surface (some properties necessary to compute the values of $\vt{S}_l^m(\theta,\phi)$ are given in \ref{app:derivative}). Figure \ref{fig:3D_modeshape} shows the normal displacement $u_r$ at the surface of the sphere ($r=a$), for the fundamental spheroidal mode ($n=1$) with $l=30$. The displacement is shown for three values of $m$ ($m=0$, $m=10$, $m=l=30$). All the modes have the same eigenfrequency because of the eigensystem degeneracy, but the modeshapes in the angular directions are quite different. These modeshapes correspond to the zonal (Fig.~\ref{fig:3D_modeshape}a), tesseral (Fig.~\ref{fig:3D_modeshape}b) and sectoral (Fig.~\ref{fig:3D_modeshape}c) patterns of the spherical harmonics~\citep{Arfken99, Kausel06}, modulated by the radial behaviour of the mode. For the angular components $u_\theta$ and $u_\phi$ (not shown here), similar patterns can be observed (some of them then involve the $\theta$-derivative of the patterns of spherical harmonics). It is noteworthy that for high values of $l$ (\textit{i.e.} with a displacement confined at the surface), the sectoral modes appear to be analogue to the so-called whispering-gallery modes, which have found many applications in optics~\citep{Knight95, Oraevsky02, Foreman15}. 

\section{Surface Acoustic Waves}
\label{sec:SAW}

In this section, the numerical model is used to reconstruct the collimation of the Rayleigh surface wave, and to interpret this phenomenon in terms of the eigenvibrations of the sphere. As an example of a multilayered sphere, the effect of a viscoelastic coating is finally investigated.

\subsection{Description of the collimating wave test case}

In this test case, an isotropic and homogeneous sphere made of viscoelastic steel is considered. The material properties are given in Table \ref{tab:materials}. The radius of the sphere is $a = \unit{25}{\milli\meter}$. A normal force $F(\theta,\phi,t)\vt{e}_r$ is applied at the surface of the sphere, with:
\begin{equation}
F(\theta,\phi,t)=f(\theta,\phi)g(t)\,.
\end{equation}
Only spheroidal modes are excited, because the excitation is limited to the radial direction.

The transient part of the force $g(t)$ is a sinus of centre frequency $f_c = \unit{1}{\mega\hertz}$ modulated over 5 cycles by a Hanning window. Besides, $f(\theta,\phi)$ is distributed along a thick line. The line source is modelled by the product of two Gaussian functions as:
\begin{equation}
f(\theta,\phi)=\mathrm{e}^{-\frac{(\theta-\theta_c)^2}{2\theta_{\sigma}^2}}\mathrm{e}^{-\frac{(\phi-\phi_c)^2}{2\phi_{\sigma}^2}}\,.
\label{eq:line_source}
\end{equation}
It is centered at the equator (\textit{i.e.} at $\theta_c=\pi/2$) and at $\phi_c=0$. The standard deviations $\theta_\sigma$ and $\phi_\sigma$ control the width of the Gaussian along the polar and the azimuthal coordinates respectively. One sets $\phi_{\sigma}=2\pi/235$ ($\approx \unit{1.5}{\degree}$) to obtain a thick line perpendicular to the equator.

According to \citet{Clorennec04}, it is possible to choose the polar aperture of the source \eqref{eq:line_source} to obtain a collimating Rayleigh wave. In that case, the propagation of the wave is diffraction-free. The Rayleigh wave propagates with a quasi-constant polar width in the direction perpendicular to the source. The collimation angle of the source is given in Ref. \cite{Clorennec04} by the formula:
\begin{equation}
\theta_{\text{COL}}=\sqrt{\frac{\pi c_R}{4af_c}},
\label{eq:collimating_angle}
\end{equation}
where $c_R \approx 0.9194c_S$ is the Rayleigh wave velocity~\citep{Eringen75a}. Here, $c_R\approx \unit{2919.8}{\meter\usk\reciprocal\second}$. It yields $\theta_{\text{COL}} \approx 0.3029$ ($\approx \unit{17.3}{\degree}$). Taking the Gaussian width as $1/e^2$, one sets $\theta_{\sigma }=0.1514$ in Eq.~\eqref{eq:line_source}. For comparison, two other cases are also considered: a longer line source ($\theta_{\sigma}= 0.2668$) and a shorter line source ($\theta_{\sigma}=0.0667$). The waves emitted by these three different sources are computed from the modal expansion given in Sec.~\ref{ssec:forced_response}. 

The source term (see Eq. \eqref{eq:FE_source}) is obtained from the coefficients $\hat{\vt{t}}_l^m$. The latter are computed with a numerical SHT analysis (see Eqs.\eqref{eq:decomposition_analysis}--\eqref{eq:GLQ}) applied to Eq.~\eqref{eq:line_source} with $l$ from $0$ to $150$. The GLQ is computed with $L=151$ Gauss points and weights (determined with the function \textit{legpts} of the Chebfun package~\citep{Driscoll14}). The number of FFT points along the azimuthal wavenumber is set to 512. It has been checked that the L2-error over the whole spherical surface between the initial source (given by Eq. \eqref{eq:line_source}) and the synthetized one (inverse SHT of $\hat{\vt{t}}_l^m$, applying Eq. \eqref{eq:switch_synthesis}) is less than 1\%. 

The length of the one-dimensional finite elements is $\Delta r = 0.003a$, which yields 2 010 dofs. The forced response is obtained with a superposition of $N=80$ eigenfrequencies. The solution is computed between $0$ and $\unit{10}{\mega\hertz}$ for 8 192 frequencies. 

\subsection{Results}

\subsubsection{Collimating, diverging and focusing waves}

\begin{figure}
\centering
\includegraphics[scale=0.25]{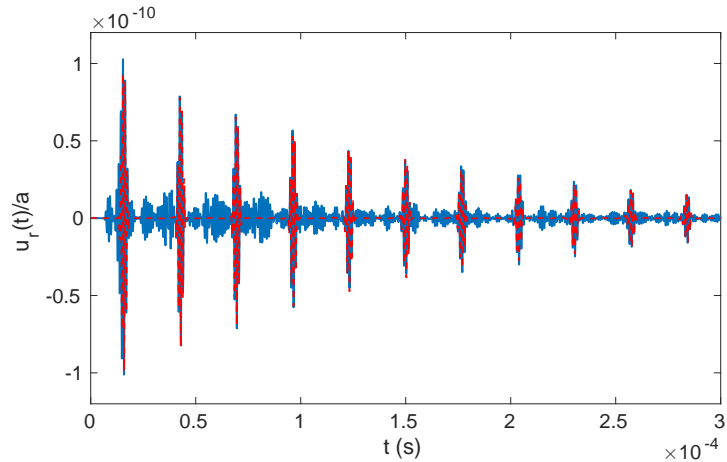}
\caption{Transient collimating signal $u_r(t)/a$ at the surface of a viscoelastic sphere ($r=a$) at point $\theta =\pi/2$, $\phi=\pi/2$. Blue curve: modal superposition using $N=80$ modes. Red dashed curve: modal superposition using the fundamental Rayleigh mode only ($N=1$). Line source: $\theta_{\sigma }=0.1514$.}
\label{fig:collimated_0_90}
\end{figure}

\begin{figure*}
\centering
\includegraphics[width=\textwidth, keepaspectratio=true]{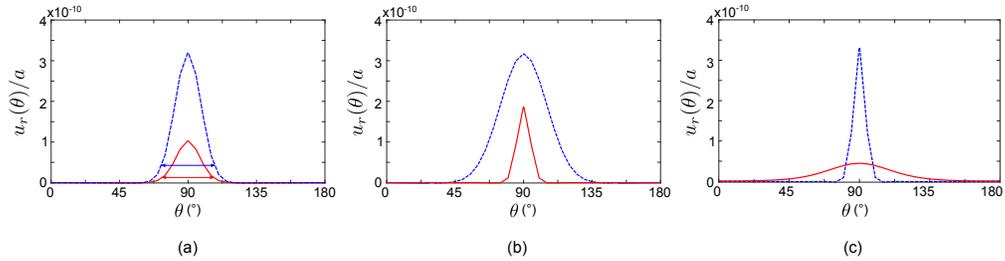}
\caption{Normal displacement $u_r(\theta)/a$ at the surface of a viscoelastic sphere ($r=a$). Blue dashed curve: at $\phi=0$ and $t=\unit{2.289}{\micro\second}$. Red solid curve: at $\phi=\pi/2$ and $t=\unit{15.38}{\micro\second}$. (a) Collimating wave ($\theta_{\sigma }=0.1514$);  (b) Focusing wave ($\theta_{\sigma}= 0.2668$); (c) Diverging wave ($\theta_{\sigma}=0.0667$).}
\label{fig:wavefront}
\end{figure*}

Figure \ref{fig:collimated_0_90} shows the transient displacement at the point $\theta=\pi/2$ (on the equator) and $\phi=\pi/2$, for the source with $\theta_{\sigma }=0.1514$. Several major peaks can be observed. These peaks correspond to the arrivals of the Rayleigh wave, either propagating counter-clockwise (\textit{e.g.} the first and the third peaks), either propagating clockwise (\textit{e.g.} the second and the fourth peaks). On this figure, the time-of-flight between the peaks is estimated to $\unit{27.01}{\micro\second}$, which agrees well with the theoretical arrival of the Rayleigh wave to do a half-trip ($\unit{26.89}{\micro\second}$).

Figure \ref{fig:wavefront} represents the normal displacement $u_r$ at the surface as a function of the polar angle $\theta$, at $\phi = 0$ (at the source position) and at $\phi=\pi/2$ (after a trip of a quarter of circumference), for the three different sources. In each case, the amplitude is lower at $\phi=\pi/2$ (red curve) than at the source position (blue dashed curve) because of viscoelastic losses (and because the source splits into waves travelling in opposite directions). 

For a source width $\theta_{\sigma }=0.1514$, the variation of the wavefront width is weak and equal to 6\% (see Fig. \ref{fig:wavefront}a). A collimating wave is then obtained, as predicted by Eq.~\eqref{eq:collimating_angle}.

Conversely, the wavefront width strongly decreases when the source is larger ($\theta_{\sigma}= 0.2668$) -- see Fig.~\ref{fig:wavefront}b. As shown in Fig.~\ref{fig:wavefront}c, when the source is shorter ($\theta_{\sigma}=0.0667$) the wavefront width increases. The propagation is not diffraction-free in these cases. Note that the diffraction reaches its maximum at $\phi=\pi/2$ because it is located right in between the source and its opposite pole. The wave in Fig.~\ref{fig:wavefront}b is called focusing wave because the wavefront focuses towards $\phi=\pi/2$ and then diverges towards the pole opposite to the source~\citep{Ishikawa03}. The so-called diverging wave (Fig.~\ref{fig:wavefront}c) diverges towards $\phi=\pi/2$ and then converges towards the pole opposite to the source. 

Some videos of the transient collimating, diverging and focusing Rayleigh waves are included in the supplementary materials of the electronic version of this paper to clearly visualize these phenomena.

\subsubsection{Modal analysis}
\label{sssec:modal_analysis}

\begin{figure}
\centering
\includegraphics[width=0.85\textwidth, keepaspectratio=true]{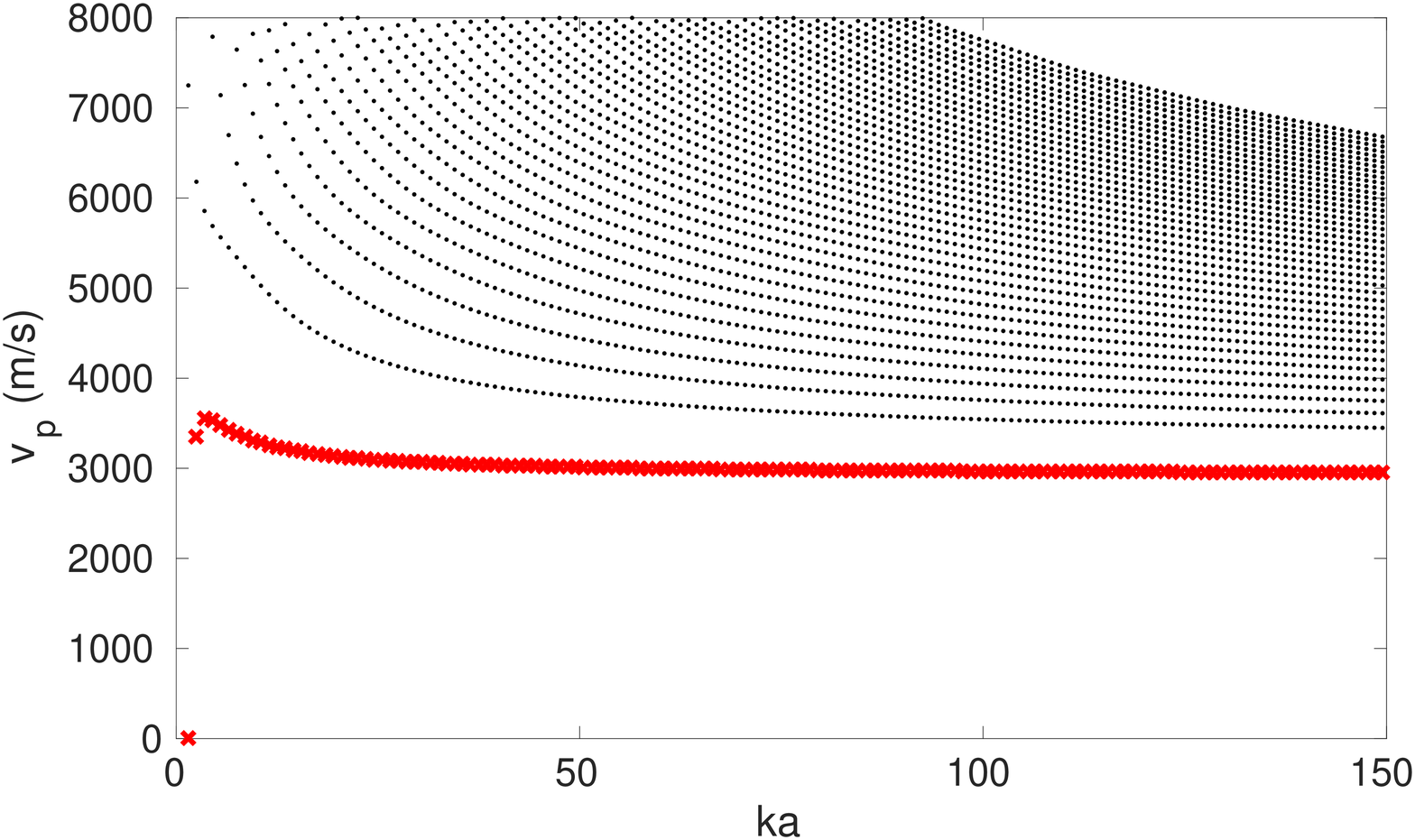}
\caption{Phase velocity of spheroidal modes for a viscoelastic steel sphere of radius $a=\unit{25}{\milli\meter}$. Red crosses: Rayleigh fundamental mode ($n=1$).c}
\label{fig:phase_velocity}
\end{figure}

As described in Sec.~\ref{sec:validation_results}, the displacement of the fundamental spheroidal mode ($n=1$) is generally confined near the surface. This mode is usually called the Rayleigh mode~\citep{Clorennec04}, because its velocity approaches asymptotically the Rayleigh wave velocity in a half-space (see Fig.~\ref{fig:phase_velocity}). Retaining only the Rayleigh mode to compute the forced response yields the red curve in Fig.~\ref{fig:collimated_0_90}, which correctly approximates the main wave packets. Higher-order modes ($n>1$) enrich the signal with the contribution of other waves which can be identified as body waves travelling inside the sphere~\citep{Eringen75a}.

\begin{figure*}
\centering
\includegraphics[width=\textwidth, keepaspectratio=true]{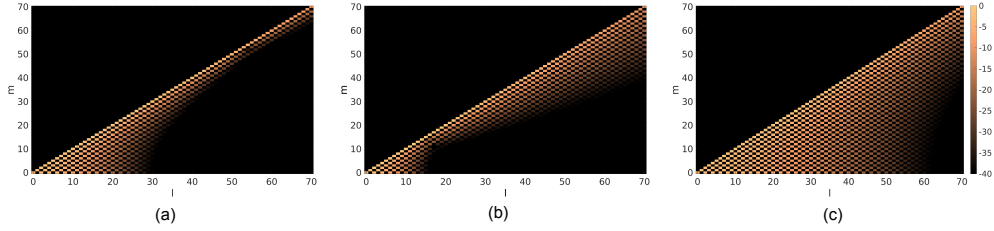}
\caption{Coefficients $10\log_{10}(|\hat{\vt{t}}_l^m/\text{max}~\hat{\vt{t}}_l^m|)$ (dB) of the force applied to obtain (a) a collimating wave ($\theta_{\sigma }=0.1514$); (b) a focusing wave ($\theta_{\sigma}= 0.2668$); (c) a diverging wave ($\theta_{\sigma}=0.0667$). These coefficients are computed from the numerical SHT analysis based on Eq.~\eqref{eq:GLQ} with $l$ from $0$ to $150$, 151 GLQ and 512 FFT points.}
\label{fig:f_l_m}
\end{figure*}

\begin{figure*}
\centering
\includegraphics[width=\textwidth, keepaspectratio=true]{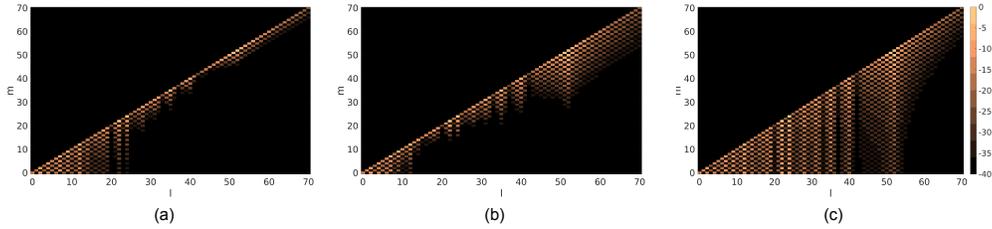}
\caption{Forced response $10\log_{10}(|\hat{u}_l^m/\text{max}~\hat{u}_l^m|)$ (dB) at the surface of a viscoelastic sphere ($r=a$) and at the centre frequency ($\overline{\omega} = 49.46$) for (a) a collimating wave ($\theta_{\sigma }=0.1514$); (b) a focusing wave ($\theta_{\sigma}= 0.2668$); (c) a diverging wave ($\theta_{\sigma}=0.0667$).}
\label{fig:u_l_m_f_c}
\end{figure*}

Interestingly the modal contributions can be further decomposed as a function of polar and azimuthal wavenumbers $l$ and $m$. Figure \ref{fig:f_l_m} displays the coefficients $|\hat{\vt{t}}_l^m|$ of the three different sources. The resulting modal responses $|\hat{u}_l^m(r=a)|$ (see Eq. \eqref{eq:FRF}) at centre frequency $\overline{\omega} = 49.46$ are shown in Fig.~\ref{fig:u_l_m_f_c}.

\begin{figure}
\centering
\includegraphics[width=0.75\textwidth, keepaspectratio=true]{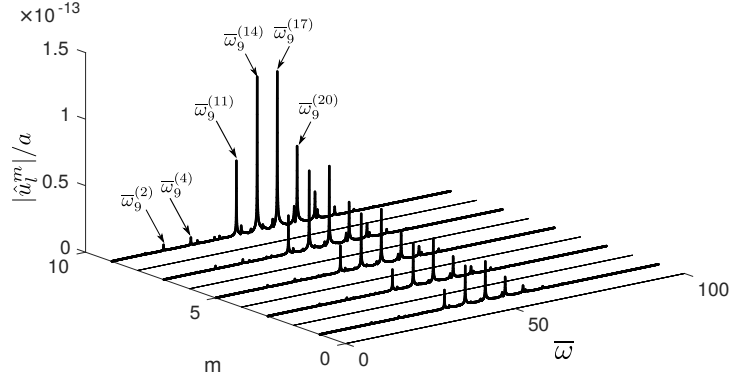}
\caption{FRF $|\hat{{u}}_l^m(\omega)|/a$ at the surface of a viscoelastic sphere ($r=a$) for $l=9$ for the collimating wave ($\theta_{\sigma }=0.1514$).}
\label{fig:frfl9}
\end{figure}

\begin{figure}
\centering
\includegraphics[width=0.75\textwidth, keepaspectratio=true]{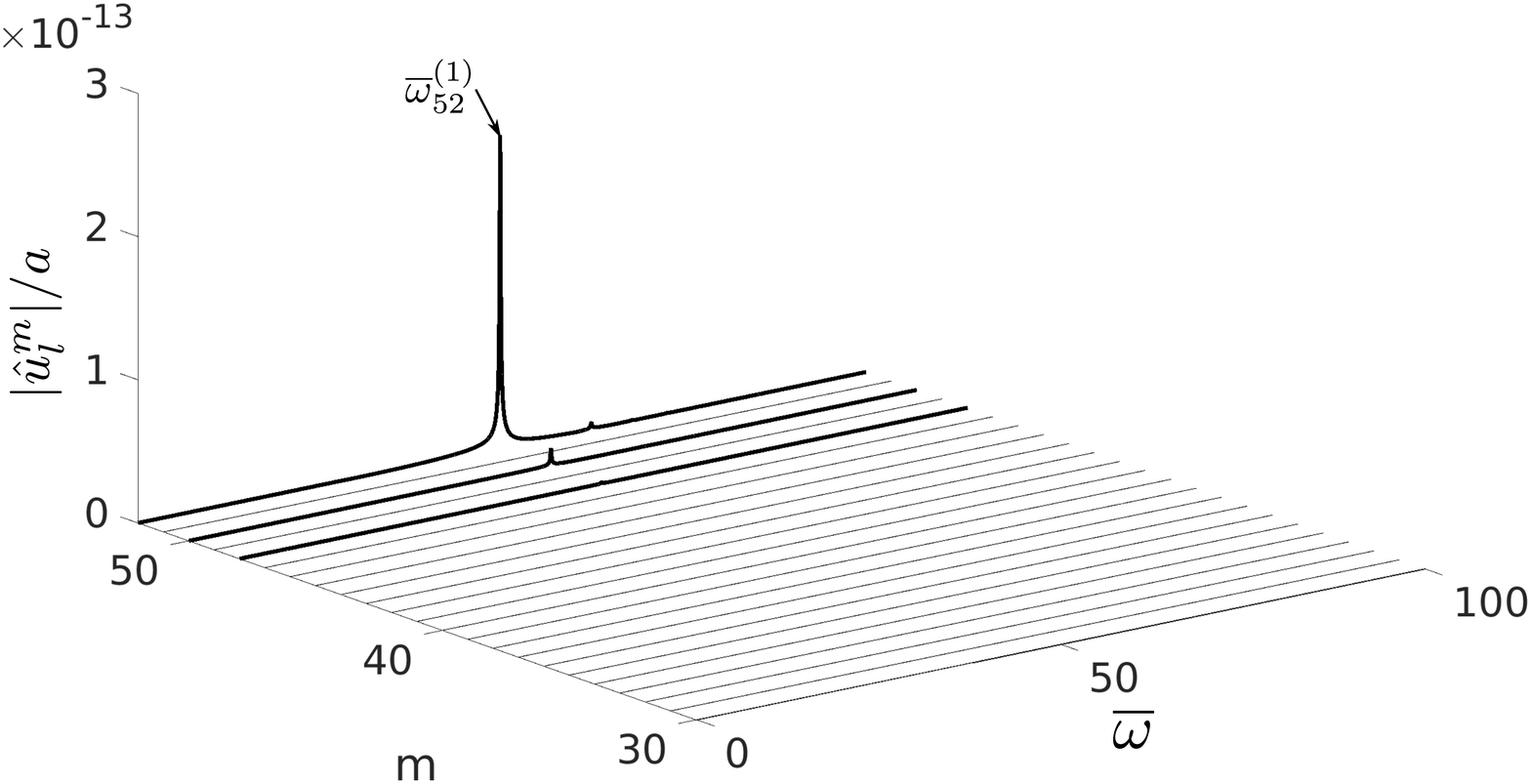}
\caption{FRF $|\hat{{u}}_l^m(\omega)|/a$ at the surface of a viscoelastic sphere ($r=a$) for $l=52$ for the collimating wave ($\theta_{\sigma}=0.1514$).}
\label{fig:frfl52_collimating}
\end{figure}

In the collimating case, Fig.~\ref{fig:f_l_m}a and Fig.~\ref{fig:u_l_m_f_c}a show that the source mostly excites sectoral modes (\textit{i.e.} with $m \approx l$), except for low values of $l$ ($l<20$) where tesseral modes ($m<l, \ m \neq 0$) also contribute. As shown in Fig.~\ref{fig:frfl9}, the FRF $|\hat{{u}}_l^m(\omega)|$ for $l=9$ exhibit several peaks which correspond to resonances of various spheroidal modes (torsional modes are not excited). It can be observed that the leading contributions come from the resonances of the 14th and the 17th spheroidal modes. The resonance of the Rayleigh mode is hence negligible. Therefore in Fig.~\ref{fig:u_l_m_f_c}a, the modes with a small polar wavenumber $l$ can be interpreted as the contribution of body waves. It can be inferred that the collimating Rayleigh wave actually corresponds to a superposition of Rayleigh sectoral modes of high wavenumber $m \approx l$, \textit{i.e.} modes confined near the surface and near the equator of the sphere. This is confirmed by Fig.~\ref{fig:frfl52_collimating}, showing the FRF $|\hat{{u}}_l^m(\omega)|$ for $l=52$.

\begin{figure}
\centering
\includegraphics[width=0.75\textwidth, keepaspectratio=true]{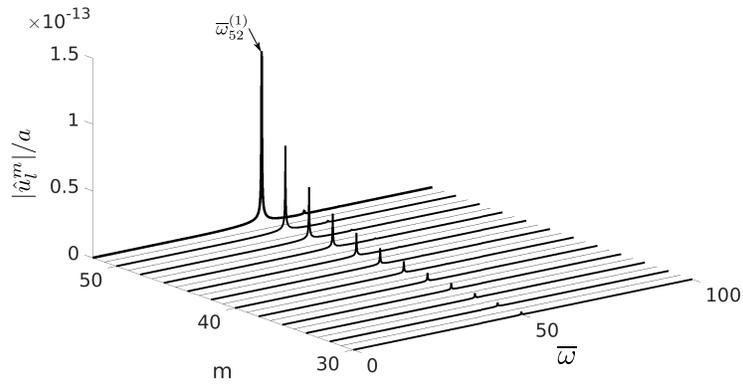}
\caption{FRF $|\hat{{u}}_l^m(\omega)|/a$ at the surface of a viscoelastic sphere ($r=a$) for $l=52$ for the diverging wave ($\theta_{\sigma}=0.0667$).}
\label{fig:frfl52}
\end{figure}

For the diverging and focusing waves, Figs.~\ref{fig:f_l_m}b-c and Figs.~\ref{fig:u_l_m_f_c}b-c show that the contribution of tesseral modes is also significant for higher polar wavenumbers $l$ ($l>30$), as opposed to the collimating case. In that case, the FRF exhibits a single resonance peak which corresponds to the Rayleigh mode for various values of $m$ (see Fig.~\ref{fig:frfl52} for the diverging case at $l=52$). Therefore, the focusing and diverging waves indeed involve the contribution of Rayleigh tesseral modes ($m \neq l$), in addition to sectoral modes. Note that in the diverging case (Fig.~\ref{fig:f_l_m}c and Fig.~\ref{fig:u_l_m_f_c}c), modes with a small polar wavenumber are more excited than in the two other cases. This is an expected result because, as already explained for the collimating source, these low-order modes represent the contribution of body waves, diffracted throughout the sphere.

As a side remark, the similarities between Figs. \ref{fig:f_l_m} and \ref{fig:u_l_m_f_c} tend to show that the type of wave (\textit{i.e.} collimating, focusing or diverging) generated by a source can be qualitatively predicted solely from the a SHT analysis of the force (Fig.~\ref{fig:f_l_m}).

\subsection{Effect of a viscoelastic coating}

\begin{figure}
\centering
 \includegraphics[width=0.85\textwidth, keepaspectratio=true]{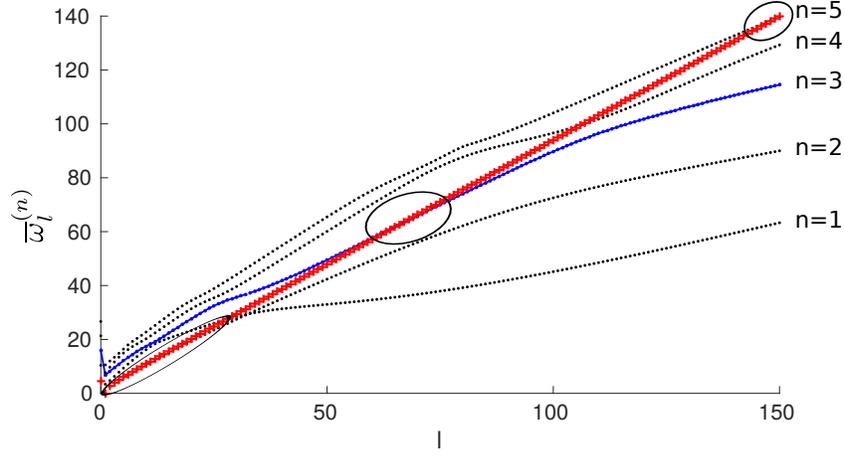}
 \caption{Non-dimensional eigenfrequencies $\overline{\omega}_l^{(n)}$ of the spheroidal modes of a viscoelastic steel sphere of radius $a=\unit{25}{\milli\meter}$. Red crosses: Rayleigh mode ($n=1$) without coating. Bullets: modes with a 1-mm coating of epoxy (in blue, the quasi-Rayleigh mode).}
 \label{fig:coating_eigenfrequencies}
\end{figure}

\begin{figure}
\centering
 \includegraphics[width=0.85\textwidth, keepaspectratio=true]{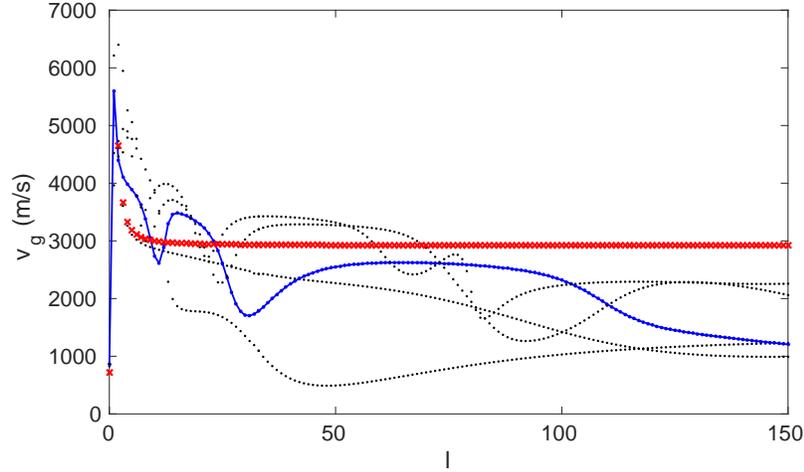}
 \caption{Group velocity of the spheroidal modes of a viscoelastic steel sphere of radius $a=\unit{25}{\milli\meter}$. Red crosses: Rayleigh mode ($n=1$) without coating. Bullets: modes with a 1-mm coating of epoxy (in blue, the quasi-Rayleigh mode).}
 \label{fig:coating_group_velocity}
\end{figure}

In this last test case, a 1-mm epoxy coating is added at the surface of the sphere (materials properties are given in Table~\ref{tab:materials}). The generation of a collimating wave at the interface between the sphere and the coating is investigated.

The eigenfrequencies of the spheroidal modes of the coated sphere are displayed in Fig.~\ref{fig:coating_eigenfrequencies} for $n \leq 5$ and the group velocity is plotted in Fig.~\ref{fig:coating_group_velocity}. For the sake of comparison, the curves corresponding to the Rayleigh mode without coating are represented with red crosses. The behaviour of the modes significantly changes with the coating. It can be observed that the modal density increases and that the modes are much more dispersive. In particular, the Rayleigh mode of the sphere is not recovered. However, for some values of $l$ the eigenfrequencies of the coated sphere almost coincide with those of the Rayleigh mode of the surface-free sphere (see the circled zone in Fig.~\ref{fig:coating_eigenfrequencies}). Group velocities can then be locally close to that of the Rayleigh wave. In Fig.~\ref{fig:coating_group_velocity}, the group velocity of the mode  identified with blue points is almost non dispersive for $50<l<100$. Its value (\unit{2620}{\meter\usk\reciprocal\second}) is ten percent lower than the Rayleigh wave velocity of the sphere. The modal attenuations (not shown here for conciseness) are also almost equal in this region. For simplicity, this mode will be called quasi-Rayleigh mode in the following.

Since the collimating wave is a superposition of Rayleigh modes (see Sec.~\ref{sssec:modal_analysis}), the source is modified to select the quasi-Rayleigh mode when it is similar to the Rayleigh mode without coating (\textit{i.e.} for $50<l<100$). The frequency bandwidth is reduced and centred on a higher frequency (the transient source $g(t)$ is a sinus of centre frequency $f_c=\unit{1.2}{\mega\hertz}$ modulated over 10 cycles). The spatial profile $f(\theta,\phi$) of the source is modified according to Eq.~\eqref{eq:collimating_angle}. The source is applied in the normal direction and at the interface between the sphere and the coating.

\begin{figure}
\centering
\includegraphics[scale=0.25]{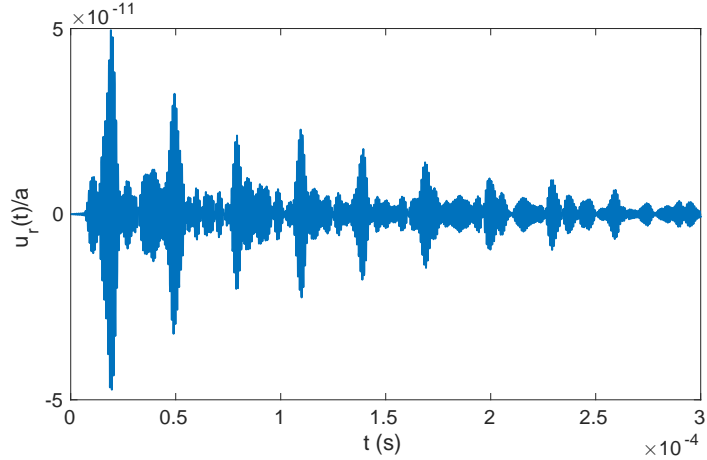}
\caption{Transient collimating signal $u_r(t)/a$ at the surface ($r=a$) of a viscoelastic sphere coated with epoxy at point $\theta =\pi/2$, $\phi=\pi/2$. Source parameter: $\theta_{\sigma }=0.1514$.}
\label{fig:collimated_coating_0_90}
\end{figure}

\begin{figure}
\centering
\includegraphics[width=0.45\textwidth, keepaspectratio=true]{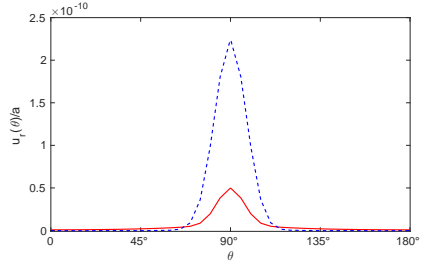}
\caption{Normal displacement $u_r(\theta)/a$ at the surface ($r=a$) of a viscoelastic sphere coated with epoxy. Blue dashed curve: at $\phi=0$ and $t=\unit{4.669}{\micro\second}$. Red solid curve: at $\phi=\pi/2$ and $t=\unit{19.14}{\micro\second}$. Source parameter: $\theta_{\sigma }=0.1514$.}
\label{fig:coating_wavefront}
\end{figure}

The forced response $u_r(t)$ at the interface and for $\theta=\pi/2$, $\phi=\pi/2$ is shown in Fig.~\ref{fig:collimated_coating_0_90}. As in the surface-free sphere (in Fig.~\ref{fig:collimated_0_90}), several major peaks are observed. The time-of-flight between the peaks is equal to \unit{30.31}{\micro\second}. It yields a velocity of \unit{2591}{\meter\usk\reciprocal\second}, which is quite close to the group velocity of the quasi-Rayleigh mode. Actually, this mode prevails in the FRF (not shown here) which confirms the modal selectivity of the chosen source. Figure \ref{fig:coating_wavefront} represents the normal displacement $u_r(\theta)$ at $\phi=0$ (blue dashed line) and $\phi=\pi/2$ (red solid line). The amplitude of the wavefront varies, but its width is nearly constant, which means that a collimating wave can be generated at the interface.

\section{Conclusion}

A semi-analytical one-dimensional finite element formulation has been proposed to compute the free and the forced responses of multi-layered spheres. The solution along the radial coordinate has been approximated with one-dimensional finite elements. Taking advantage of both vector and tensor spherical harmonics orthogonality, the appropriate choice of test function has led to independent governing equations for each couple of angular wavenumbers. A fully analytical description of the angular behaviour of the displacement fields as well as a general formulation suitable for any interpolating technique has been eventually obtained. The formulation yields a linear eigenvalue problem which is simple and fast to solve. The eigensolutions give both the spheroidal and torsional modes. The accuracy of the model has been checked by comparison with literature results for a homogeneous sphere.

The vibration modes have been superposed to reconstruct surface acoustic waves phenomena in the sphere. A collimating Rayleigh wave has then been recovered numerically. The modal analysis of such a wave, based on the resonances of the sphere, has shown that the collimating Rayleigh wave corresponds to a superposition of the fundamental spheroidal modes with a displacement confined at the equator of the sphere: the so-called Rayleigh modes, of sectoral type, with a high polar wavenumber. When the sphere is coated with a thin elastic layer, the numerical results have shown that the Rayleigh mode behaviour is approximately recovered in a limited frequency range. This allows generating a collimating wave at the interface of the sphere and the coating.

Further works are currently in progress to model and study the influence of an infinite embedding medium on the free vibrations and surface wave propagation in spherical structures.

\section*{Acknowledgement}

This work was founded by Région Pays de la Loire (Project SMOg).

\clearpage

\onecolumn

\appendix

\section{Properties of spherical harmonics}

\subsection{Scalar orthogonality relationships}
\label{app:orthogonality}

First, let us recall that the normalized spherical harmonics form an orthonormal basis, such that~\citep{Arfken99}:
\begin{equation}
\int_0^\pi \int_0^{2\pi}Y_k^{p*}Y_l^m\mathrm{d}\phi\sin\theta\mathrm{d}\theta=\delta_{kl}\delta_{mp}\,.
\label{eq:orthoYY}
\end{equation}
Integrating by parts and using the Legendre equation \eqref{eq:legendre_equation}, one can show that~\citep{Kausel06}:
\begin{equation}
\int_0^\pi \int_0^{2\pi}\left[\frac{\partial Y_k^{p*}}{\partial\theta}\frac{\partial Y_l^m}{\partial\theta}+\frac{1}{\sin^2\theta}\frac{\partial Y_k^{p*}}{\partial\phi}\frac{\partial Y_l^m}{\partial\phi}\right]\mathrm{d}\phi\sin\theta\mathrm{d}\theta=\overline{l}\delta_{kl}\delta_{mp}\,.
\label{eq:orthoY'Y'}
\end{equation}

Furthermore, it can be readily shown that:
\begin{equation}
\int_0^\pi \int_0^{2\pi} \left[ \frac{\mathrm{d}Y_k^{p*}}{\mathrm{d}\theta}Y_l ^m+Y_k^{p*}\frac{\mathrm{d}Y_l^m}{\mathrm{d}\theta}\right]\mathrm{d}\phi\sin\theta\mathrm{d}\theta=0\,.
\label{eq:orthoY'Y}
\end{equation}
These three scalar relationships yields the orthogonality of vector spherical harmonics, expressed by Eq.~\eqref{eq:orth_Slm}, and are also useful to evaluate some components of the stiffness and mass matrices.

Others relationships are necessary to evaluate the integral of $\delta\vt{\epsilon}^{\mathrm{T}}\vt{\sigma}$ in Eq.~\eqref{eq:variational_formulation}. These relations are \cite{Martinec00}:
\begin{multline}
\int_0^\pi \int_0^{2\pi}\left[
\left(\frac{\partial^2 Y_k^{p*}}{\partial \theta^2}-\cot\theta\frac{\partial Y_k^{p*}}{\partial\theta}-\frac{1}{\sin^2\theta}\frac{\partial^2 Y_k^{p*}}{\partial \phi^2}\right)\right.
 \left(\frac{\partial^2 Y_l^m}{\partial \theta^2}-\cot\theta\frac{\partial Y_l^m}{\partial\theta}-\frac{1}{\sin^2\theta}\frac{\partial^2 Y_l^m}{\partial \phi^2}\right)\\
 \left.+4\frac{\partial}{\partial\theta}\left(\frac{1}{\sin\theta}\frac{\partial Y_k^{p*}}{\partial \phi}\right)\frac{\partial}{\partial\theta}\left(\frac{1}{\sin\theta}\frac{\partial Y_l^m}{\partial \phi}\right)\right]\mathrm{d}\phi\sin\theta\mathrm{d}\theta=(l-1)\overline{l}(l+2)\delta_{kl}\delta_{mp},
\label{eq:ortho_tensor1}
\end{multline}
\begin{multline}
\int_0^\pi \int_0^{2\pi}\left[
-\frac{\partial}{\partial\theta}\left(\frac{1}{\sin\theta}\frac{\partial Y_k^{p*}}{\partial \phi}\right)\left(\frac{\partial^2 Y_l^m}{\partial \theta^2}-\cot\theta\frac{\partial Y_l^m}{\partial\theta}-\frac{1}{\sin^2\theta}\frac{\partial^2 Y_l^m}{\partial \phi^2}\right)\right.\\
\left.+\frac{\partial}{\partial\theta}\left(\frac{1}{\sin\theta}\frac{\partial Y_l^m}{\partial \phi}\right)\left(\frac{\partial^2 Y_k^{p*}}{\partial \theta^2}-\cot\theta\frac{\partial Y_k^{p*}}{\partial\theta}-\frac{1}{\sin^2\theta}\frac{\partial^2 Y_k^{p*}}{\partial \phi^2}\right)\right]\mathrm{d}\phi\sin\theta\mathrm{d}\theta=0.
\label{eq:ortho_tensor2}
\end{multline}
The above results, which are given in a scalar form in this paper for the sake of simplicity, must actually be derived from the orthogonality properties of tensor spherical harmonics. This derivation is more mathematically involved than for vector spherical harmonics~\citep{James76}. These tensorial properties can be found under a compact dyadic form in the work of \citet{Martinec00}.

\subsection{Derivative with respect to the polar angle}
\label{app:derivative}

To evaluate the value of the matrix $\vt{S}_l^m$ at any angular coordinates, it is necessary to compute the derivative $\frac{\partial  Y_l^m}{\partial \theta}$. Convenient formulas are given by \citet{Bosch00} to avoid singular values at poles. These formulas have been slightly modified to be consistent with the normalization chosen in this paper, based on the guidelines of Ref. \cite[Appendix A]{Bosch00}, denoting $\overline{P}_l^m(\cos\theta)=N_l^mP_l^m(\cos\theta)$.

The polar derivative of a spherical harmonic can be obtained using the following recurrence formula (for $m \geq 0$):
\begin{equation}
2\frac{\partial Y_l^m(\theta,\phi)}{\partial \theta}=\left(\sqrt{(l-m)(l+m+1)}\overline{P}_l^{m+1}(\cos\theta)\right.\left.-\sqrt{(l+m)(l-m+1)}\overline{P}_l^{m-1}(\cos\theta)\right)\frac{\mathrm{e}^{\mathrm{j}m\phi}}{\sqrt{2\pi}}\,.
\label{eq:derivative_spherical_harmonics}
\end{equation}
For $m < 0$, the derivative can be readily obtained using the equality $Y_l^{-m}=(-1)^m Y_l^m$. The cases of $m=0$ or $m=l$ are specific and the derivatives are given by:
\begin{align}
\frac{\partial Y_0^0(\theta,\phi)}{\partial \theta}&=0\,,\label{eq:derivative_0_0}\\
\frac{\partial Y_l^0(\theta,\phi)}{\partial \theta}&=\sqrt{\frac{\overline{l}}{4\pi}}\overline{P}_l^1(\cos\theta)\,, \label{eq:derivative_l_0}\\
\frac{\partial Y_l^l(\theta,\phi)}{\partial \theta}&=-\sqrt{\frac{l}{4\pi}}\overline{P}_l^{l-1}(\cos\theta)\mathrm{e}^{\mathrm{j}l\phi}\,.\label{eq:derivative_l_l}
\end{align}

At poles ($\theta=0$ or $\theta=\pi$), for $m=0$ the spherical harmonics do not depend on $\phi$. From Eq.~\eqref{eq:norm_spherical_harmonic} and using the properties $P_l^0(1)=1$ and $P_l^0(-1)=(-1)^l$, one gets~\citep{Arfken99}:
\begin{align}
Y_l^0(0,\phi)=\sqrt{\frac{2l+1}{4\pi}}\,, \\
Y_l^0(\pi,\phi)=(-1)^l\sqrt{\frac{2l+1}{4\pi}}\,.
\end{align}
For $m>0$, the azimuthal coordinate is undetermined and the spherical harmonics must hence vanish at poles~\citep{Arfken99}. 

Similarly, the derivative must vanish at poles for $m \neq 0$. For $m=0$, using the property $\overline{P}_l^1(\pm 1)=0$ into Eq.~\eqref{eq:derivative_l_0} enables to show that:
\begin{equation}
 \frac{\partial Y_l^0(0,\phi)}{\partial \theta}=\frac{\partial Y_l^0(\pi,\phi)}{\partial \theta}=0\,.
\end{equation}

\section{Example: calculation of the component $K_{22}$ of the stiffness matrix}
\label{app:example}

In the following, let us detail for the sake of clarity the computation of the second diagonal component of the stiffness matrix $\vt{K}$, denoted $K_{22}$. The latter is derived from the integral of $\delta\vt{\epsilon}^{\mathrm{T}}\vt{\sigma}$ in Eq.~\eqref{eq:variational_formulation}, where the integrand can be explicitely written: 
\begin{multline}
\delta\vt{\epsilon}^{\mathrm{T}}\vt{\sigma}=\sum_{l \geq 0} \sum_{|m|\leq l} \left[\frac{\partial \delta\hat{\vt{u}}^{\mathrm{T}}}{\partial r}\vt{S}_k^{p*}\vt{L}_r^{\mathrm{T}}\vt{C}\vt{L}_r\vt{S}_l^m \frac{\partial \hat{\vt{u}}_l^m}{\partial r}\right.\\
+\left(\frac{1}{r}\frac{\partial \delta\hat{\vt{u}}^{\mathrm{T}}}{\partial r}\vt{S}_k^{p*}\vt{L}_r^{\mathrm{T}}+\frac{\delta\hat{\vt{u}}^{\mathrm{T}}}{r^2}\frac{\partial \vt{S}_k^{p*}}{\partial \theta}\vt{L}_{\theta}^{\mathrm{T}}+\frac{\delta\hat{\vt{u}}^{\mathrm{T}}}{r^2\sin\theta}\frac{\partial \vt{S}_k^{p*}}{\partial \phi}\vt{L}_{\phi}^{\mathrm{T}}+\frac{\delta\hat{\vt{u}}^{\mathrm{T}}}{r^2}\vt{S}_k^{p*}\vt{L}_{1}^{\mathrm{T}}+\cot\theta\frac{\delta\hat{\vt{u}}^{\mathrm{T}}}{r^2}\vt{S}_k^{p*}\vt{L}_{2}^{\mathrm{T}}\right)\vt{C}\vt{A}_l^m\hat{\vt{u}}_l^m\\ 
+\left.\delta\hat{\vt{u}}^{\mathrm{T}}\vt{A}_k^{p*\mathrm{T}}\vt{C}\vt{S}_l^m\vt{L}_r\frac{1}{r}\frac{\partial \hat{\vt{u}}_l^m}{\partial r}\right],
\end{multline}
where $\vt{A}_l^m=\left(\vt{L}_\theta\frac{\partial}{\partial \theta}\vt{S}_l^m+\frac{1}{\sin\theta}\vt{L}_{\phi}\frac{\partial}{\partial \phi}\vt{S}_l^m+\vt{L}_1\vt{S}_l^m+\cot\theta\vt{L}_2\vt{S}_l^m \right)$.

Hence, there is 25 matrices to compute using Eqs.~\eqref{eq:C_matrix}, \eqref{eq:L_operator}, \eqref{eq:Slm}. In the specific case of the component $K_{22}$, it yields on one finite element:
\begin{multline}
K_{22}=\sum_{l \geq 0} \sum_{|m|\leq l} \int_0^\pi \int_0^{2\pi} \left[\int\delta\hat{\vt{u}}^{e\mathrm{T}} \left(\left(\frac{\partial^2Y_k^{p*}}{\partial\theta^2}+\cot\theta \frac{\partial Y_k^{p*}}{\partial\theta}+ \frac{1}{\sin^2\theta}\frac{\partial^2 Y_k^{p*}}{\partial \phi^2}\right)\right.\right.\\
\times \left(\frac{\partial^2 Y_l^m}{\partial\theta^2}+ \cot\theta \frac{\partial Y_l^m}{\partial\theta}+ \frac{1}{\sin^2\theta}\frac{\partial^2 Y_l^m}{\theta\partial \phi^2}\right)C_{23}\\  
+\left\lbrace 2\left(\cot\theta \frac{\partial Y_k^{p*}}{\partial\theta}+\frac{1}{\sin^2\theta} \frac{\partial^2 Y_k^{p*}}{\partial \phi^2}\right) \left(\cot\theta \frac{\partial Y_l^m}{\partial\theta}+ \frac{1}{\sin^2\theta}\frac{\partial^2 Y_l^m}{\partial \phi^2}\right) + 2 \frac{\partial^2Y_k^{p*}}{\partial\theta^2} \frac{\partial^2 Y_l^m}{\partial\theta^2}\right.\\
 +\left.\left.\left(\frac{\partial}{\partial\theta}\left(\frac{1}{\sin\theta}\frac{\partial Y_k^{p*}}{\partial \phi}\right)-  \frac{\cot\theta}{\sin\theta	}\frac{\partial Y_k^{p*}}{\partial\phi} +\frac{1}{\sin\theta}\frac{\partial^2Y_k^{p*}}{\partial\phi\partial\theta} 
  \right)\left(\frac{\partial}{\partial\theta}\left(\frac{1}{\sin\theta}\frac{\partial Y_l^m}{\partial \phi}\right)-  
\frac{\cot\theta}{\sin\theta}  \frac{\partial Y_l^m}{\partial\phi}+\frac{1}{\sin\theta}\frac{\partial^2 Y_l^m}{\partial\theta\partial\phi}\right)\right\rbrace C_{44}\right)\hat{\vt{u}}_l^{me}\mathrm{d}r\\
+\left.\int \left(\delta\hat{\vt{u}}^{e\mathrm{T}}-r\frac{\partial \delta\hat{\vt{u}}^{e\mathrm{T}}}{\partial r}	\right) \left(\frac{1}{\sin^2\theta}\frac{\partial Y_k^{p*}}{\partial\phi}\frac{\partial Y_l^m}{\partial\phi}+\frac{\partial Y_k^{p*}}{\partial\theta}\frac{\partial Y_l^m}{\partial\theta}\right)
\left(\hat{\vt{u}}^{e}-r\frac{\partial \hat{\vt{u}}_l^{me}}{\partial r} \right)C_{55}\mathrm{d}r \right] \mathrm{d}\phi \sin \theta \mathrm{d}\theta\,.
\end{multline}

It can be immediatly noticed that the terms factored by $C_{55}$ involves the orthogonality relationship \eqref{eq:orthoY'Y'}. Noticing that $-\frac{\cot\theta}{\sin\theta}  \frac{\partial Y_l^m}{\partial\phi}+\frac{1}{\sin\theta}\frac{\partial^2 Y_l^m}{\partial\theta\partial\phi}=\frac{\partial}{\partial\theta}\left(\frac{1}{\sin\theta}\frac{\partial Y_l^m}{\partial \phi}\right)  
$, and using the Legendre equation \eqref{eq:legendre_equation}, the term factored by $C_{44}$ can be rewritten as:
\begin{multline}
\overline{k}~\overline{l}Y_k^{p*}Y_l^{m}+\left(\frac{\partial^2 Y_k^{p*}}{\partial \theta^2}-\cot\theta\frac{\partial Y_k^{p*}}{\partial\theta}-\frac{1}{\sin^2\theta}\frac{\partial^2 Y_k^{p*}}{\partial \phi^2}\right)\left(\frac{\partial^2 Y_l^m}{\partial \theta^2}-\cot\theta\frac{\partial Y_l^m}{\partial\theta}-\frac{1}{\sin^2\theta}\frac{\partial^2 Y_l^m}{\partial \phi^2}\right)\\
+4\frac{\partial}{\partial\theta}\left(\frac{1}{\sin\theta}\frac{\partial Y_k^{p*}}{\partial \phi}\right)\frac{\partial}{\partial\theta}\left(\frac{1}{\sin\theta}\frac{\partial Y_l^m}{\partial \phi}\right)\,,
\end{multline} 
where $\overline{k}=k(k+1)$. This term involves orthogonality relationships \eqref{eq:orthoYY} and \eqref{eq:ortho_tensor1}. Finally, the two terms factored by $C_{23}$ can be replaced using the Legendre equation \eqref{eq:legendre_equation} and further simplified using the orthogonality relationship \eqref{eq:orthoYY}. This yields:
\begin{equation}
 K_{22}=\int\delta\hat{\vt{u}}^{e\mathrm{T}}\left(\overline{l}^2 C_{23}+2\overline{l}(\overline{l}-1)C_{44}\right)\hat{\vt{u}}_l^{me}\mathrm{d}r 
 +\int \left(\delta\hat{\vt{u}}^{e\mathrm{T}}-r\frac{\partial \delta\hat{\vt{u}}^{e\mathrm{T}}}{\partial r}	\right) \overline{l}C_{55} \left(\hat{\vt{u}}^{e}-r\frac{\partial\hat{\vt{u}}^{e}}{\partial r}	\right)\mathrm{d}r.
\end{equation}
Using the finite element interpolation \eqref{eq:FE_discretization}, one finally recasts the second diagonal component of the elementary matrices of Eqs.~\eqref{eq:elementary_matrixK1}--\eqref{eq:elementary_matrixK3}.

\clearpage

\bibliographystyle{unsrtnat}
\bibliography{bib_surface_free_spherical}

\end{document}